\documentclass[aps,prc,preprint,showpacs]{revtex4}
\usepackage{graphicx}
\usepackage{textcomp}

\begin{document}

\title{Strong decay of low-lying {\boldmath $S_{11}$} and {\boldmath $D_{13}$} nucleon resonances to
 pseudoscalar mesons and octet baryons}

\author{C. S. An}
\email[]{chunsheng.an@cea.fr}
\affiliation{Institut de Recherche sur les
lois Fondamentales de l'Univers, DSM/Irfu, CEA/Saclay, F-91191 Gif-sur-Yvette, France}

\author{B. Saghai}
\email[]{bijan.saghai@cea.fr}
\affiliation{Institut de Recherche sur les
lois Fondamentales de l'Univers, DSM/Irfu, CEA/Saclay, F-91191 Gif-sur-Yvette, France}

\date{\today}

\begin{abstract}

Partial decay widths of lowest lying nucleon resonances $S_{11}(1535)$, $S_{11}(1650)$, $D_{13}(1520)$ and $D_{13}(1700)$ 
to the pseudoscalar mesons and octet baryons are studied within a chiral constituent quark model. 
Effects of the configurations mixing between the states $|N^{2}_{8}P_{M}\rangle$ and $|N^{4}_{8}P_{M}\rangle$ 
are considered, taking into account $SU(6)\otimes O(3)$ breaking effects. 
In addition, possible contributions of the strangeness components in the $S_{11}$ resonances are investigated. 
Experimental data for the partial decay widths of the $S_{11}$ and $D_{13}$ resonances are well reproduced. 
Predictions for coupling constants of the four nucleon resonances to pseudoscalar mesons and octet baryons, crucial issues in the 
photo- and hadron-induced meson production reactions, are reported.
Contributions from five-quark components in the $S_{11}$ resonances are found crucial in reproducing the partial widths.

\end{abstract}

\maketitle

\section{Introduction}
\label{intro}
Production of mesons with hidden or open strangeness via electromagnetic or hadronic probes,
in the baryon resonance energy range, is subject to extensive experimental and theoretical investigations. 
In this realm, partial decay widths of resonances to meson-baryon final states, as well as the relevant
coupling constants are crucial, but not well enough known~\cite{Nakamura:2010zzi}, ingredients in our 
understanding of the reaction mechanisms, and also of the nature of those resonances.

Phenomenological approaches, dealing with the above ingredients, arise mainly from two families of 
formalisms: effective Lagrangians based on meson-baryon degrees of freedom~\cite{Kaiser:1995cy,Riska:2000gd,Penner:2002ma,Penner:2002md,Inoue:2001ip,Aznauryan:2003zg,Liu:2005pm,Chiang:2001pw,Chiang:2004ye,Nakayama:2004ek,Nakayama:2005ts,Nakayama:2008tg,Vrana:1999nt,Matsuyama:2006rp,JuliaDiaz:2007kz,Durand:2008es,Cao:2008st,Shyam:2003uz,Shyam:2007iz,Shklyar:2006xw,Arndt:2005dg,Anisovich:2005tf,Sarantsev:2005tg,Kaptari:2008ae,Ceci:2006jj,
Geng:2008cv,Xie:2010md,Bruns:2010sv,Gamermann:2011mq}
and QCD based/inspired models
~\cite{Koniuk:1979vy,Capstick:1992th,Capstick:1993kb,Capstick:1998uh,Capstick:2000qj,Capstick:2004tb,Kiswandhi:2003ca,Golli:2011jk,Kim:1997ym,Yoshimoto:1999dr,Zhu:1998qw,Zhu:1998am,Jido:1997yk,Hyodo:2008xr,Goity:2005rg,Jayalath:2011uc}.

Among the low-lying nucleon excitations, the $S_{11}(1535)$ resonance plays a special role due to its large 
$\eta N$ decay width~\cite{Nakamura:2010zzi}, though its mass is very close to the threshold of the decay. 
Moreover, in the $KY$ production reactions the importance of the $S_{11}(1650)$ is well established. 
For the two other first orbitally excited (quark model prediction) nucleon resonances, $D_{13}(1520)$ and 
$D_{13}(1700)$, the couplings to the pseudoscalar meson and octet baryons seem to be rather weak, but the first one 
is known to intervene significantly in the polarization asymmetries.

The observables of interest in this paper are partial decay widths. Experimental values are available~\cite{Nakamura:2010zzi}
for all four resonances' decay to $\pi N$ and $\eta N$ final states, as well as for the $S_{11}(1650)$ and $D_{13}(1700)$
resonances to $K \Lambda$, though with rather large uncertainties. 
However, in spite of extensive studies mentioned above, to our knowledge no single formalism has 
reproduced {\it simultaneously} those partial widths. The only exception here is a very recent comprehensive 
study~\cite{Jayalath:2011uc} based on the $1/N_C$ expansion approach.
Besides the fact that a large number of investigations concentrate on the $S_{11}$ resonances, recent copious 
photoproduction data have not yet been fully exploited by sophisticated coupled-channels phenomenological approaches. 
The main motivation of the present work is then to study those partial decay widths within a QCD inspired formalism,
and shed light on the structure of those baryons.

The theoretical frame of the present work is based on a chiral constituent quark model ($\chi$CQM), 
complemented with the $SU(6)\otimes O(3)$ symmetry breaking effects.
The outcomes of those formalisms are compared to the known~\cite{Nakamura:2010zzi} partial decay widths 
of the above mentioned resonances. This approach gives satisfactory results for the $D_{13}$ resonances,
but misses partly the data for $S_{11}$.

Attempting to cure the observed theory / experiment discrepancies, the $\chi$CQM is subsequently
complemented with including contributions from higher Fock-components, namely, five-quark configurations. 
Actually, several authors~\cite{Li:2005jn,JuliaDiaz:2006av,An:2005cj,Li:2006nm,Li:2005jb,An:2008xk,An:2008tz,An:2010wb}, 
have shown that contributions from the five-quark components are quite significant in describing the 
properties of baryons and their electromagnetic and strong decays, especially contributions from the 
$qqqq\bar{q} \to  M(\gamma)+qqq$ transitions. 
For recent reviews on five-quark components in baryons, see Refs.~\cite{Zou:2009zz,Zou:2010tc,riska}.

The extended $\chi$CQM allows reproducing the known partial decay widths for both $S_{11}$ 
resonances. 
Following the successful results obtained for low-lying baryon resonances, we put 
forward predictions for the coupling constants of those resonances to seven meson-baryon final states, i.e. 
$\pi^{0} p$, $\pi^{+} n$, $\eta p$, $K^{+} \Lambda$, $K^{0}\Sigma^{+}$, $K^{+} \Sigma^{0}$, $\eta^{\prime} p$.

The present manuscript is organized in the following way: in section~\ref{theo}, we present the theoretical 
formalism which includes the wave functions, strong decays and the resulting transition amplitudes for the
$S_{11}(1535)$, $S_{11}(1650)$, $D_{13}(1520)$ and $D_{13}(1700)$ to the pseudoscalar 
mesons and octet baryons. Numerical results are given in section~\ref{num}, and finally
section~\ref{con} contains summary and conclusions.
%
%
\section{Theoretical Formalism}
\label{theo}
In section~\ref{wfm}, we present the wave functions of the nucleon resonances $S_{11}(1535)$, $S_{11}(1650)$, 
$D_{13}(1520)$ and $D_{13}(1700)$.
Section~\ref{form} embodies a brief review of the formalism for the strong decay of the baryon
resonances to meson-baryon in a $\chi$CQM, where we derive transition coupling amplitudes for the above four 
nucleon resonances to the $\pi N$, $\eta N$, $K\Lambda$, $K\Sigma$ and $\eta^{\prime}N$ channels.
%
%
\subsection{Wave functions}
\label{wfm}

In the $\chi$CQM, complemented with five-quark components, a baryon is a superposition 
of three- and five-quark mixture and the wave function can be written as
\begin{equation}
 |B\rangle=A_{3}|qqq\rangle+A_{5}|qqqq\bar{q}\rangle\,,
\end{equation}
with $A_{3}$ and $A_{5}$ the probability amplitudes for the corresponding $qqq$ and $qqqq\bar{q}$ states, 
respectively.

For the three-quark components, we employ the wave functions in traditional three-quark $\chi$CQM. 
In the $SU(6)\otimes O(3)$ conserved case, the general form for the wave functions of the octet
baryons, $N(^{2}_{8}P_{M})_{S^{-}}$ and $N(^{4}_{8}P_{M})_{S^{-}}$ states, can be expressed as
\begin{eqnarray}
|B(^{2}_{8}S_{S})_{\frac{1}{2}^{+}},S_{z}\rangle&=&\frac{1}{\sqrt{2}}(|B\rangle_{\lambda}|\frac{1}{2},s_{z}\rangle_{\lambda}
+|B\rangle_{\rho}|\frac{1}{2},s_{z}\rangle_{\rho})\varphi_{000}^{s}(\vec{\lambda},\vec{\rho})\,,\\ 
 |N(^{2}_{8}P_{M})_{S^{-}},S_{z}\rangle&=&\frac{1}{2}\sum_{m,s_{z}}C^{SS_{z}}_{1m,\frac{1}{2}s_{z}}
 [(|N\rangle_{\rho}|\frac{1}{2},s_{z}\rangle_{\lambda}+|N\rangle_{\lambda}|\frac{1}{2},s_{z}\rangle_{\rho})
 \varphi^{\rho}_{11m}(\vec{\lambda},\vec{\rho})\nonumber\\
&&
+(|N\rangle_{\rho}|\frac{1}{2},s_{z}\rangle_{\rho}-|N\rangle_{\lambda}|\frac{1}{2},s_{z}\rangle_{\lambda})
 \varphi^{\lambda}_{11m}(\vec{\lambda},\vec{\rho})]\,, 
\label{suwfb} \\
|N(^{4}_{8}P_{M})_{S^{-}},S_{z}\rangle&=&\frac{1}{\sqrt{2}}\sum_{m,s_{z}}C^{SS_{z}}_{1m,\frac{3}{2}s_{z}}
[|N\rangle_{\rho}|\frac{3}{2},s_{z}\rangle
\varphi^{\rho}_{11m}(\vec{\lambda},\vec{\rho})
+|N\rangle_{\lambda}|\frac{3}{2},s_{z}\rangle
\varphi^{\lambda}_{11m}(\vec{\lambda},\vec{\rho})]\,,
\label{suwfc}
\end{eqnarray}
where $|B\rangle_{\rho(\lambda)}$ denotes the mixed symmetric flavor wave function of the 
three-quark system for the corresponding baryon. 
$|\frac{1}{2},s_{z}\rangle_{\rho(\lambda)}$ and $|\frac{3}{2},s_{z}\rangle$ are the mixed 
symmetric and symmetric spin wave functions of the three-quark system, respectively.
$\varphi_{Nlm}(\vec{\lambda},\vec{\rho})$ is the harmonic oscillator basis orbital wave 
function for the three quarks with the subscripts $Nlm$ being the corresponding quantum 
numbers. 
Finally, $C^{SS_{z}}_{1m,ss_{z}}$ are the Clebsch-Gordan coefficients for the coupling of the
orbital and spin of the three-quark system to form a baryon state with spin $S$ and z-component 
$S_{z}$. 
The explicit forms for all of the above flavor, spin, and orbital wave functions can be found 
in~\cite{An:2010wb}.

Taking into account the breakdown of $SU(6)\otimes O(3)$ symmetry due to either the 
color-magnetic~\cite{De Rujula:1975ge} or flavor-magnetic~\cite{Glozman:1995fu} hyperfine 
interactions between the quarks, one can express the wave functions of the $S_{11}$ and $D_{13}$ 
resonances in terms of the given $N(^{2}_{8}P_{M})_{S^{-}}$ and $N(^{4}_{8}P_{M})_{S^{-}}$ wave
functions, Eqs.~(\ref{suwfb}) and~(\ref{suwfc}) , by introducing the configuration mixing angles 
$\theta _{S}$ and $\theta_{D}$
\begin{eqnarray}
\pmatrix{ |S_{11}(1535)\rangle\cr |S_{11}(1650)\rangle\cr}&=&
\pmatrix{ cos\theta_{S}  &  -sin\theta_{S} \cr sin\theta_{S}   &
cos\theta_{S} \cr} \pmatrix{
|N(^{2}_{8}P_{M})_{\frac{1}{2}^{-}}\rangle \cr
|N(^{4}_{8}P_{M})_{\frac{1}{2}^{-}}\rangle \cr}\,,
\label{ThetaS}
\end{eqnarray}

\begin{eqnarray}
\pmatrix{ |D_{13}(1520)\rangle\cr |D_{13}(1700)\rangle\cr}&=&
\pmatrix{ cos\theta_{D}  &  -sin\theta_{D} \cr sin\theta_{D}   &
cos\theta_{D} \cr} \pmatrix{
|N(^{2}_{8}P_{M})_{\frac{3}{2}^{-}}\rangle \cr
|N(^{4}_{8}P_{M})_{\frac{3}{2}^{-}}\rangle \cr}.
\end{eqnarray}

For the octet baryons, other than the lowest lying $S_{11}$ and $D_{13}$, the configuration mixing 
effects are not so significant. 
So, for those baryons we take the wave functions within the exact $SU(6)\otimes O(3)$ symmetry.

For the five-quark components of $S_{11}(1535)$, we use the wave functions given in Ref.~\cite{An:2008tz},

\begin{eqnarray}
\psi_{t,s}&=&\sum_{a,b,c}\sum_{Y,y,T_z,t_z}\sum_{S_z,s_z}
C^{[1^4]}_{[31]_a[211]_a} C^{[31]_a}_{[F]_b [S]_c}
[F]_{b,Y,T_z} [S]_{c,S_z}
[211;C]_a
(Y,T,T_z,y,\bar t,t_z|1,1/2,t)\nonumber\\
&&(S,S_z,1/2,s_z|1/2,s)\bar\chi_{y,t_z}\bar\xi_{s_z}\varphi_{[5]}\, .
\label{wfc}
\end{eqnarray}
In fact, this general wave function is appropriate for the five-quark components in all 
the low-lying nucleon resonances with $S^{p}=\frac{1}{2}^{-}$, albeit with different
probabilities for five-quark components.

As reported in Ref.~\cite{An:2008tz}, there are 5 different flavor-spin configurations which may 
form five-quark components in the resonances with negative parity. 
If the hyperfine interaction between the quarks is assumed to depend on flavor and spin, the energy of 
the second and third configurations should be about 80 MeV and 200 MeV higher than the first configuration, 
respectively. 
Since $S_{11}(1535)$ and $S_{11}(1650)$ are the first two orbital excitations of the nucleon with 
spin $1/2$, the configurations with low energies, namely the first two five-quark configurations should
be the most appropriate ones to form higher Fock components in those two resonances. 
Moreover, the contribution of the second five-quark configuration is very similar
to that of the first one, because of the same flavor structure, which rules out the five-quark components 
with light quark and anti-quark pairs in the $S_{11}$ resonances. 
Actually, the transition elements between all of the 5 five-quark configurations and the octet baryons differ 
just by constant factors. 
Therefore, the contributions from all the 5 configurations are similar, albeit with appropriate probability 
amplitudes. 
Consequently, the first configuration is enough for us to study the strong decays of $S_{11}(1535)$ and 
$S_{11}(1650)$. 
Then the wave functions for the five-quark components in $S_{11}(1535)$ and $S_{11}(1650)$ reduce
to the following form:
\begin{equation}
 \psi_{5q}=\sum_{abc}C^{[1^4]}_{[31]_{a}
[211]_{a}}C^{[31]_{a}}_{[211]_{b}[22]_{c}}[4]_{X}[211]_{F}(b)[22]_{S}(c)[211]_{C}(a)
\bar \chi_{s_{z}}\varphi(\{\vec{\xi}_i\})\,,
\end{equation}
the explicit form of which is given in Ref.~\cite{An:2008xk}. 

Following Eq.~(\ref{ThetaS}), the introduction of five-quark wave functions leads to
\begin{eqnarray}
 |S_{11}(1535)\rangle&=&A_{3}\left[cos\theta_{S}|N(^{2}_{8}P_{M})_{\frac{1}{2}^{-}}\rangle-sin\theta_{S}
|N(^{4}_{8}P_{M})_{\frac{1}{2}^{-}}\rangle \right]+A_{5}\psi_{5q},\\
|S_{11}(1650)\rangle&=&A_{3}^{'}\left[sin\theta_{S}|N(^{2}_{8}P_{M})_{\frac{1}{2}^{-}}\rangle+cos\theta_{S}
|N(^{4}_{8}P_{M})_{\frac{1}{2}^{-}}\rangle\right]+A_{5}^{\prime}\psi_{5q}\,.
\label{wfc}
\end{eqnarray}
The probability amplitude for the five-quark component in a baryon can be related to the coupling 
$_{5q}\langle\hat{V}_{cou} \rangle_{3q}$ between the $qqq$ and $qqqq\bar{q}$ configurations in the 
corresponding baryon 
\begin{eqnarray}
A_{5q}=\frac{_{5q}\langle\hat{V}_{cou} \rangle_{3q}}{M-E_{5}},
\label{amp5}
\end{eqnarray}
with $E_{5}$ the energy of the five-quark component. 
Given that the resonances considered here have negative parity, all of the quarks and anti-quark in 
the five-quark system should be in their ground states. 
Hence, we can take $\hat{V}_{cou}$ to be of the following form:
\begin{equation}
 \hat{V}_{cou}=3V(r_{34})\frac{\hat{\sigma}_{3}\cdot\vec{p}_{3}}{2m_{3}}\chi_{00}^{45}C_{00}^{45}
F_{00}^{45}\varphi_{00}(\vec{p}_{4}-\vec{p}_{5})b_{4}^{\dagger}(\vec{p}_{4})d_{5}^{\dagger}(\vec{p}_{5})\,,
\end{equation}
where $\chi_{00}^{45}$, $C_{00}^{45}$, $F_{00}^{45}$ and $\varphi_{00}(\vec{p}_{4}-\vec{p}_{5})$ 
denote the spin, flavor, color and orbital singlets of the quark and anti-quark pair, respectively. 
$b_{4}^{\dagger}(\vec{p}_{4})$ and $d_{5}^{\dagger}(\vec{p}_{5})$ are the creation operators for a
quark and anti-quark pair with momentum $\vec{p}_{4}$ and $\vec{p}_{5}$, respectively. $V(r_{34})$ 
is the coupling potential which depends on the relative coordinate $|\vec{r}_{3}-\vec{r}_{4}|$. 
Then we obtain
\begin{equation}
\frac{\langle\psi_{5q}|\hat{V}_{cou}|N(^{2}_{8}P_{M})_{\frac{1}{2}^{-}}\rangle}
{\langle\psi_{5q}|\hat{V}_{cou}|N(^{4}_{8}P_{M})_{\frac{1}{2}^{-}}\rangle}=-2,
\end{equation}
and
\begin{equation}
\frac{A_{5q}^{\prime}}{A_{5q}} = \frac{sin\theta_{S}-\frac{1}{2}cos\theta_{S}}{cos\theta_{S}+\frac{1}{2}sin\theta_{S}}
\frac{M_{S_{11}(1535)}-E_{5}}{M_{S_{11}(1650)}-E_{5}}.
\label{ampratio}
\end{equation}

Here we would like to emphasize that the considered $D_{13}$ resonances are not relevant for five-quark components
issues.
Actually, all of the quarks and anti-quark should be in their ground states (lowest energy) to form
the negative parity. Then the spin configuration of four-quark subsystem is limited to be $[31]_{S}$, 
for which the total spin of the four-quark subsystems is $S=1$, in order to combine with the anti-quark 
to form the required total spin $3/2$.
For the configurations with spin $[31]_{S}$, the flavor-spin overlap factors between such five-quark configurations 
and the $D_{13}$ states vanish. 
Therefore, the probabilities for these five-quark components in the $D_{13}$ resonances are $0$. 
Some additional five-quark configurations, other than those given in Ref.~\cite{An:2008tz}, could also be considered,
for instance, the configurations with the anti-quark orbitally excited ($l_{\bar{q}}=2,4\cdots$), the ones in which the 
four-quark subsystem with spin symmetry $[4]_{S}$ ($S_{4}=2$), or the ones given in Ref.~\cite{An:2005cj} with the four 
quark subsystem orbital symmetry $[31]_{X}$ and orbital momentum $L_{4}=2,4\cdots$.
However, all those configurations have very high energies, far away from the lowest lying $D_{13}$ resonances masses.

Finally, we do not consider the five-quark components in the ground states of octet baryons in this manuscript, 
because on the one hand their probabilities in the baryons are very small~\cite{JuliaDiaz:2006av,Li:2007hn}, 
and on the other hand their contributions to electromagnetic and strong decays of nucleon resonances are 
negligible~\cite{An:2008xk}. 
Actually, the five-quark configurations in the ground states of octet baryons cannot transit to three-quark 
components of the first orbitally excited baryon resonances due to the vanishing flavor-spin overlap factors.

\subsection{Formalism for strong decay}
\label{form}

It is well known that the pseudoscalar meson-quark coupling, in the tree level approximation, takes the form
\begin{equation}
 H_M=\sum_{j}\frac{g^{q}_{A}}{2f_{M}}\bar{\psi}_{j}\gamma^{j}_{\mu}\gamma_{5}^{j}\psi_{j}\partial^{\mu}\phi_{M}\,,
\label{cou}
\end{equation}
where $\psi_{j}$ and $\phi_{M}$ are the quark and pseudoscalar fields, respectively, and $g^{q}_{A}$ is the axial 
coupling constant for the constituent quarks, the value of which is in the range 
$0.7-1.26$~\cite{Goity:1998jr,Riska:2000gd,Lahde:2002fe}. 
$f_{M}$ denotes the decay constant of the corresponding meson; the empirical values for the decay constants of 
$\pi$, $K$, $\eta$ and $\eta^{\prime}$ are $f_{\pi}=93$ MeV, $f_{K}=113$ MeV, $f_{\eta}=1.2f_{\pi}$, 
$f_{\eta^{\prime}}=-0.58f_{\pi}$.

In the framework of non-relativistic $qqq$ quark model, the coupling, Eq.~(\ref{cou}), takes the following form:
\begin{equation}
H^{NR(3)}_{M}=\sum_{j}\frac{g^{q}_{A}}{2f_{M}}(\frac{\omega_{M}}{E_{f}+M_{f}}\sigma\cdot\vec{P}_{f}+
\frac{\omega_{M}}{E_{i}+M_{i}}\sigma\cdot\vec{P}_{i}-\sigma\cdot\vec{k}_{M}
+\frac{\omega_{M}}{2\mu}\sigma\cdot\vec{p}_{j})X^{j}_{M}\exp\{-i\vec{k}_{M}\cdot\vec{r}_{j}\}\,.
\label{op3}
\end{equation}
Here, $\vec{k}_{M}$ and $\omega_{M}$ are the three momentum and energy of the final meson, $\vec{P}_{i(f)}$ and 
$M_{i(f)}$ denote the mass and three momentum of the initial (final) baryon, $\vec{p}_{j}$ and $\vec{r}_{j}$ the 
three momentum and coordinate of the $j^{th}$ quark, and $\mu$ is the reduced mass of the initial and final $j^{th}$ 
quark which emits the meson. Finally, $X^{j}_{M}$ is the flavor operator for emission of the meson from the corresponding 
$j^{th}$ quark, given by following expressions:
\begin{eqnarray}
& X^{j}_{\pi^{0}}=\lambda_{3}^{j}, ~X^{j}_{\pi^{\pm}}=\mp\frac{1}{\sqrt{2}}(\lambda_{1}^{j}\mp\lambda_{2}^{j}),\nonumber\\
&
X^{j}_{K^{\pm}}=\mp\frac{1}{\sqrt{2}}(\lambda_{4}^{j}\mp\lambda_{5}^{j}),~X^{j}_{K^{0}}=
\mp\frac{1}{\sqrt{2}}(\lambda_{6}^{j}\mp\lambda_{7}^{j}),\\
&
X^{j}_{\eta}=cos\theta\lambda_{8}^{j}-sin\theta\sqrt{\frac{2}{3}}\mathcal{I},
~X^{j}_{\eta^\prime}=sin\theta\lambda_{8}^{j}+cos\theta\sqrt{\frac{2}{3}}\mathcal{I}\nonumber\,,
\end{eqnarray}
where $\lambda^{j}_{i}$ are the $SU(3)$ Gell-Mann matrices, and $\mathcal{I}$ the
unit operator in the $SU(3)$ flavor space. $\theta$ denotes the mixing angle between
$\eta_{1}$ and $\eta_{8}$, leading to the physical $\eta$ and $\eta ^\prime$
\begin{eqnarray}
 \eta  &=& \eta _8 cos \theta -  \eta _1 sin \theta \,,\\
 \eta ^\prime &=& \eta _8 sin \theta +  \eta _1 cos \theta \,,
\label{eq:etamix}
\end{eqnarray}
it takes the value $\theta=-23$\textdegree~\cite{Gobbi:1993au}.

\begin{figure}[t]
\begin{center}
\includegraphics[scale=0.6]{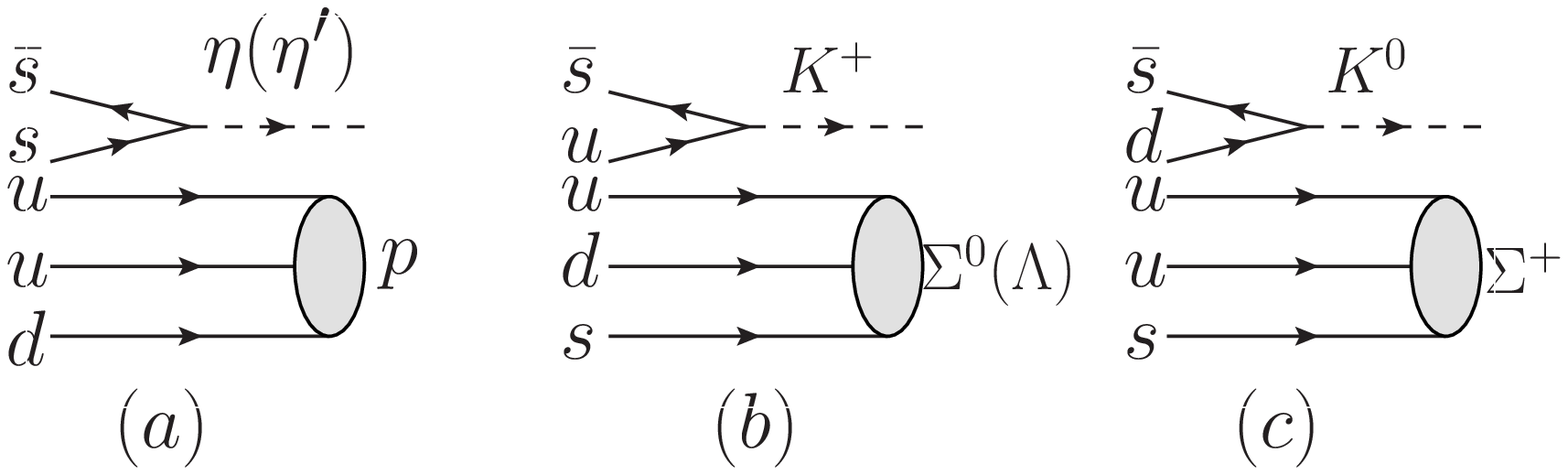}
\end{center}
\caption{(Color online)\footnotesize Strangeness component transit in the $S_{11}$ resonances
 to $\eta p$ or $\eta^\prime p$ (a), $K^+ \Lambda$ or $K^+ \Sigma ^0$ (b) and $K^0 \Sigma^+$ (c).
\label{fig}}
\end{figure}

Taking into account the five-quark components in the resonances, we have to calculate the transition coupling 
amplitudes for $qqqq\bar{q}\to qqq+M$. The reduced form of the coupling in Eq.~(\ref{cou}) reads
\begin{equation}
 H_{M}^{NR(5)}=\sum_{j}\frac{g^{q}_{A}}{2f_{M}}C_{XFSC}^{j}(m_{i}+m_{f})
 \bar{\chi}^{\dagger}_{z}\pmatrix{ 1&0\cr 0&1
\cr}\chi_{z}^{j}X_{M}^{j}\exp\{-i\vec{k}_{M}\cdot\vec{r}_{j}\}\,,
\label{op5}
\end{equation}
where $m_{i}$ and $m_{f}$ denote the constituent masses of the quark and anti-quark which combine to form a 
pseudoscalar meson, $C_{XFSC}^{j}$ denotes the overlap between the three-quark configuration of the final baryon 
and the residual orbital-flavor-spin-color configuration of the three-quark system that is left in the initial 
$qqqq\bar{q}$ after the combination of the $j^{th}$ quark with the anti-quark into a final meson. 
The transitions $qqqs\bar{s}\to B+M$ scheme is shown in Fig.~\ref{fig}. where three quarks of the five-quark 
system go as spectators to form the final three-quark baryon, and the fourth quark gets combined with the strange anti-quark 
to form a meson: $K$, $\eta$ or $\eta^{\prime}$.

Then, the transition coupling amplitude for a resonance to a pseudoscalar meson and a octet baryon is obtained 
by calculating the following matrix element:
\begin{equation}
T^{MB}=\langle
B(^{2}_{8}S_{S})_{\frac{1}{2}^{+}}|(H_{M}^{NR(3)}+H_{M}^{(5)})|N^{*}\rangle
\equiv T^{MB}_{3}+T^{MB}_{5},
\end{equation}
the resulting transition coupling amplitudes $T^{MB}_{3}$ and $T^{MB}_{5}$ for the $S_{11}$ and $D_{13}$ 
resonances to $\pi^{0}p$, $\pi^{+}n$, $\eta p$, $K^{+}\Lambda$, $K^{0}\Sigma^{+}$, $K^{+}\Sigma^{0}$ and 
$\eta^{'}p$ channels are shown in Tables~\ref{a3qsd} and~\ref{a5q}, respectively.
\begin{table}[t]
\caption{\footnotesize Transition coupling amplitudes $T^{MB}_{3}$ for the low-lying $S_{11}$ and $D_{13}$
resonances to meson-baryon final states. 
Note that the full amplitudes are obtained by multiplying each term by the following expressions:
$\frac{g^{q}_{A}}{2f_{M}}\omega_{3}[(a_{M}-\frac{b_{M}}{3})\frac{k_{M}^{2}}{\omega_{3}^{2}}-3b_{M}]
\exp\{-\frac{k^{2}_{M}}{6\omega_{3}^{2}}\}$ for $S_{11}$ and 
$\frac{g^{q}_{A}}{2f_{M}}(a_{M}-\frac{b_{M}}{3})\frac{k_{M}^{2}}{\omega_{3}}
\exp\{-\frac{k^{2}_{M}}{6\omega_{3}^{2}}\}$ for $D_{13}$ resonances. 
Here, $\omega_{3}$ is the harmonic oscillator parameter for the three-quark components,
$a_{M}=1+\frac{\omega_{M}}{E_{f}+M_{f}}$ and $b_{M}=\frac{\omega_{M}}{2\mu}$.}
\begin{tabular}{lcccc}
\hline \hline
             & $S_{11}(1535)$ &&&   $S_{11}(1650)$ \\
\hline

$\pi^{0}p$ &  $\frac{\sqrt{2}}{9}(2cos\theta_{S}-sin\theta_{S})$ &&&
$\frac{\sqrt{2}}{9}(2sin\theta_{S}+cos\theta_{S})$
\\

$\pi^{+}n$ &  $-\frac{2}{9}(2cos\theta_{S}-sin\theta_{S})$ &&&
$-\frac{2}{9}(2sin\theta_{S}+cos\theta_{S})$
\\

$\eta p$  &
$\frac{\sqrt{2}}{3}(cos\theta_{S}+sin\theta_{S})(\frac{1}{\sqrt{3}}cos\theta-\sqrt{\frac{2}{3}}sin\theta)$
&&&
$\frac{\sqrt{2}}{3}(sin\theta_{S}-cos\theta_{S})(\frac{1}{\sqrt{3}}cos\theta-\sqrt{\frac{2}{3}}sin\theta)$
 \\

$K^{0}\Lambda$ & $-\frac{1}{\sqrt{6}}cos\theta_{S}$  &&&
$-\frac{1}{\sqrt{6}}sin\theta_{S}$
\\

$K^{0}\Sigma^{+}$ & $-\frac{1}{9}(cos\theta_{S}+4sin\theta_{S})$ &&&
$-\frac{1}{9}(sin\theta_{S}-4cos\theta_{S})$
\\

$K^{+}\Sigma^{0}$ &
$-\frac{1}{9\sqrt{2}}(cos\theta_{S}+4sin\theta_{S})$ &&&
$-\frac{1}{9\sqrt{2}}(sin\theta_{S}-4cos\theta_{S})$
\\

$\eta^{\prime}p$ &
$\frac{\sqrt{2}}{3}(cos\theta_{S}+sin\theta_{S})(\frac{1}{\sqrt{3}}sin\theta+\sqrt{\frac{2}{3}}cos\theta)$
&&&
$\frac{\sqrt{2}}{3}(sin\theta_{S}-cos\theta_{S})(\frac{1}{\sqrt{3}}sin\theta+\sqrt{\frac{2}{3}}cos\theta)$

\\

\hline \hline
             & $D_{13}(1520)$ &&&   $D_{13}(1700)$ \\
\hline

$\pi^{0}p$ &
$-\frac{2}{9}(2cos\theta_{D}-\frac{1}{\sqrt{10}}sin\theta_{D})$ &&&
$-\frac{2}{9}(2sin\theta_{D}+\frac{1}{\sqrt{10}}cos\theta_{D})$
\\

$\pi^{+}n$ &
$\frac{2\sqrt{2}}{9}(2cos\theta_{D}-\frac{1}{\sqrt{10}}sin\theta_{D})$
&&&
$\frac{2\sqrt{2}}{9}(2sin\theta_{D}+\frac{1}{\sqrt{10}}cos\theta_{D})$
\\

$\eta p$  &
$-\frac{2}{3}(cos\theta_{D}+\frac{1}{\sqrt{10}}sin\theta_{D})(\frac{1}{\sqrt{3}}cos\theta-\sqrt{\frac{2}{3}}sin\theta)$
&&&
$-\frac{2}{3}(sin\theta_{D}-\frac{1}{\sqrt{10}}cos\theta_{D})(\frac{1}{\sqrt{3}}cos\theta-\sqrt{\frac{2}{3}}sin\theta)$
 \\

$K^{0}\Lambda$ & $\frac{1}{\sqrt{3}}cos\theta_{D}$  &&&
$\frac{1}{\sqrt{3}}sin\theta_{D}$
\\

$K^{0}\Sigma^{+}$ &
$\frac{1}{9}(\sqrt{2}cos\theta_{D}+\frac{4}{\sqrt{5}}sin\theta_{D})$
&&&
$\frac{1}{9}(\sqrt{2}sin\theta_{D}-\frac{}{\sqrt{5}}4cos\theta_{D})$
\\

$K^{+}\Sigma^{0}$ &
$-\frac{1}{9\sqrt{2}}(\sqrt{2}cos\theta_{D}+\frac{4}{\sqrt{5}}sin\theta_{D})$
&&&
$-\frac{1}{9\sqrt{2}}(\sqrt{2}sin\theta_{D}-\frac{4}{\sqrt{5}}cos\theta_{D})$
\\

$\eta^{\prime}p$ &
$-\frac{2}{3}(cos\theta_{D}+\frac{1}{\sqrt{10}}sin\theta_{D})(\frac{1}{\sqrt{3}}sin\theta+\sqrt{\frac{2}{3}}cos\theta)$
&&&
$-\frac{2}{3}(sin\theta_{D}-\frac{1}{\sqrt{10}}cos\theta_{D})(\frac{1}{\sqrt{3}}sin\theta+\sqrt{\frac{2}{3}}cos\theta)$
\\
\hline \hline
\end{tabular}
\label{a3qsd}
\end{table}
%
Notice that (Table~\ref{a3qsd}), within the exact $SU(6)\otimes O(3)$ symmetry, the matrix elements for 
transition $N(^{4}_{8}P_{M})_{S^{-}}\to K\Lambda$ vanish, and hence the decay widths of 
$S_{11}(1650)$ and $D_{13}(1700)$ to $K\Lambda$ are null. 
Moreover, Table~\ref{a5q}, the transition elements for $5q\to MB$ do not vanish when $k_{M}=0$, 
and it may enhance or depress the transitions $S_{11}\to MB$ significantly near the meson-baryon threshold.
Finally, the strangeness component does not transit to $\pi^{0}p$,  since the matrix element of the flavor operator 
$X^{j}_{\pi^{0}}$ between the $s\bar{s}$ pair is $0$.

\begin{table}[ht!]
\caption{\footnotesize Transition coupling amplitudes $T^{MB}_{5}$. Note that the full amplitudes are obtained by
multiplying each term by the following expression:
$\frac{g^{q}_{A}}{2f_{M}}C_{35}\exp\{-\frac{3k^{2}_{M}}{20\omega_{5}^{2}}\}$,  with $C_{35}$ related to the harmonic 
oscillator parameter for the three- and five-quark components as
$C_{35}=(\frac{2\omega_{3}\omega_{5}}{\omega_{3}^{2}+\omega_{5}^{2}})^{3}$.}
\begin{tabular}{ccccccccccccc}
\hline \hline
              $\pi^{0}p$ && $\pi^{+}n$ &&  $\eta p$ &&  $K^{+}\Lambda$ && $K^{0}\Sigma^{+}$ && $K^{+}\Sigma^{0}$ &&  $\eta^{\prime}p$\\
\hline

$0$ && $0$ &&$\frac{2}{\sqrt{3}}m_{s}(2cos\theta+\sqrt{2}sin\theta)$ &&
$\frac{1}{\sqrt{3}}(m+m_{s})$ && $\sqrt{2}(m+m_{s})$ &&
-$(m+m_{s})$ &&
$\frac{2}{\sqrt{3}}m_{s}(2sin\theta-\sqrt{2}cos\theta)$
\\
\hline \hline
\end{tabular}
\label{a5q}
\end{table}
%

To obtain the relevant expressions for partial decay widths, we take the Lagrangian for $N^{*}MB$ coupling
in hadronic level to be of the following form:
\begin{eqnarray}
\mathcal{L}_{S_{11}BM}&=&-ig_{S_{11}BM}\bar{\psi}_{B}\phi_{M}\psi_{S_{11}}+h.c.,\\
\mathcal{L}_{D_{13}BM}&=&\frac{1}{m_{M}} g_{D_{13}BM} \bar{\psi}_{B}\partial_{\mu}\phi_{M}\psi_{D_{13}}^{\mu}+h.c.,
\label{lag}
\end{eqnarray}
where $\bar{\psi}_{B}$ and $\psi_{S_{11}}$ denote the Dirac spinor fields for the final baryon and the $S_{11}$ 
resonances, respectively, and $\phi_{M}$ is the scalar field for the final meson. 

For the $D_{13}$ resonances, with spin $3/2$, we employ the Rarita-Schwinger vector-spinor fields
$\psi_{D_{13}}^{\mu}$~\cite{Rarita:1941mf,Nath:1971wp}, which are defined as
\begin{eqnarray}
\psi_{D_{13}}^{\mu}(S_{z})=\sum_{ms}C^{\frac{3}{2}S_{z}}_{1m,\frac{1}{2}s}
\epsilon^{\mu}_{m}u_{s}. 
\end{eqnarray} 

 One can directly obtain the transition coupling amplitudes for $N^{*}\to MB$ in the hadronic level
using the Lagrangian, Eq.~(\ref{lag}). Then, the coupling constants $g_{N^{*}MB}$ are extracted by comparing 
the transition coupling amplitudes $T^{MB}$ in the quark model to those in the hadronic model.

With the resulting coupling constants, the strong decay widths for the $S_{11}$ and $D_{13}$ resonances to 
the pseudoscalar meson and octet baryon read
\begin{eqnarray}
\Gamma_{S_{11}\to MB}&=&\frac{1}{4\pi}g_{S_{11}MB}^{2}\frac{E_{f}+M_{f}}{M_{i}}|\vec{k}_{M}|\,,
\label{widthS}
\end{eqnarray}
\begin{eqnarray}
\Gamma_{D_{13}\to MB}&=&\frac{1}{12\pi}\frac{1}{m_{M}^{2}} g_{D_{13}MB}^{2}
\frac{E_{f}-M_{f}}{M_{i}}|\vec{k}_{M}|^{3}\,.
\label{widthD}
\end{eqnarray}

Note that in the center of mass frame of the initial resonance, $\vec{P}_{i}=0$, $\vec{k}_{M}$ 
and $E_{f}$ can be related to the masses of the initial and final hadrons as
\begin{eqnarray}
|\vec{k}_{M}|&=&|\vec{P}_{f}|=\frac{\sqrt{[M^{2}_{i}-(M_{f}+m_{M})^{2}][M^{2}_{i}-(M_{f}
-m_{M})^{2}]}}{2M_{i}}\,,\\
\nonumber\\
E_{f}&=&\sqrt{|\vec{k}_{M}|^{2}+M_{f}^{2}}=\frac{M_{i}^{2}-m_{M}^{2}+M_{f}^{2}}{2M_{i}}.
\label{kMEf}
\end{eqnarray}

For decay channels with thresholds above the mass of the initial resonance, 
off-shell effects are taken into account by  putting $|\vec{k}_{M}|=0$ and introducing the form factor~\cite{Penner:2002ma}
\begin{equation}
F=\frac{\Lambda^{4}}{\Lambda^{4}+(q_{N^{*}}^{2}-M_{N^{*}}^{2})^{2}}\,,
\end{equation}
with the cutoff parameter $\Lambda=1$ GeV, and $q_{N^{*}}$ the threshold of the corresponding channel. 
In fact, this form factor affects mainly the 
$N^* \to \eta^{\prime}N$ process, since thresholds for all other channels are below or slightly above the masses of 
the four resonances.
%
%
\section{Numerical Results}
\label{num}
In this section our results for partial decay widths $\Gamma_{N^* \to MB}$ and
coupling constants $g_{N^* MB}$ are reported for the four investigated resonances, with 
$MB \equiv \pi^{0}p$, $\pi^{+}n$, $\eta p$, $K^{+}\Lambda$, $K^{0}\Sigma^{+}$, $K^{+}\Sigma^{0}$ 
and $\eta^{\prime}p$.

The starting point, section ~\ref{num3q}, is the standard $\chi$CQM. 
Then, in section~\ref{num3qb} we introduce $SU(6)\otimes O(3)$ breaking  and finally, 
in section~\ref{numtt}, the five-quark components are embodied for the $S_{11}$ resonances.

For the partial decay widths, we compare our results to the experimental values reported in
PDG~\cite{Nakamura:2010zzi}, and produce predictions for yet unmeasured channels.
\subsection{Pure $qqq$ configuration and exact $SU(6)\otimes O(3)$ symmetry}
\label{num3q}
Within this simplest configuration, there are three input parameters: quarks' masses and harmonic 
oscillator parameter.

For the constituent quarks' masses, we use the traditional $qqq$ quark model values~\cite{Glozman:1995fu,Koniuk:1979vy,Capstick:2000qj}, namely, 
$m \equiv  m_{u}=m_{d}=340$ MeV and $m_{s}=460$ MeV.

The scale of the oscillator parameter, $\omega_{3}$, can be inferred from the
empirical radius of the proton via $\omega_{3}=1/\sqrt{\langle r^{2}\rangle}$,
which leads to $\omega_{3}\simeq250$ MeV, for $\sqrt{\langle r^{2}\rangle} \simeq 1$ fm.
However, since the photon couples to $u$ and $d$ quarks through $\rho$ and $\omega$ mesons, 
the measured proton charge radius may reflect partly the vector meson propagator~\cite{weisebook}. 
Moreover, pion cloud have some influence on the measured proton charge radius. Consequently, 
the intrinsic size of the proton still has some model dependence, and hence, the oscillator parameter
$\omega_{3}$ might deviate from 250 MeV, 
within the range $100-400$ MeV~\cite{Li:2006nm,JuliaDiaz:2006av,Koniuk:1979vy,Capstick:2000qj}.

\begin{figure}[t]
\begin{center}
\includegraphics[scale=0.23]{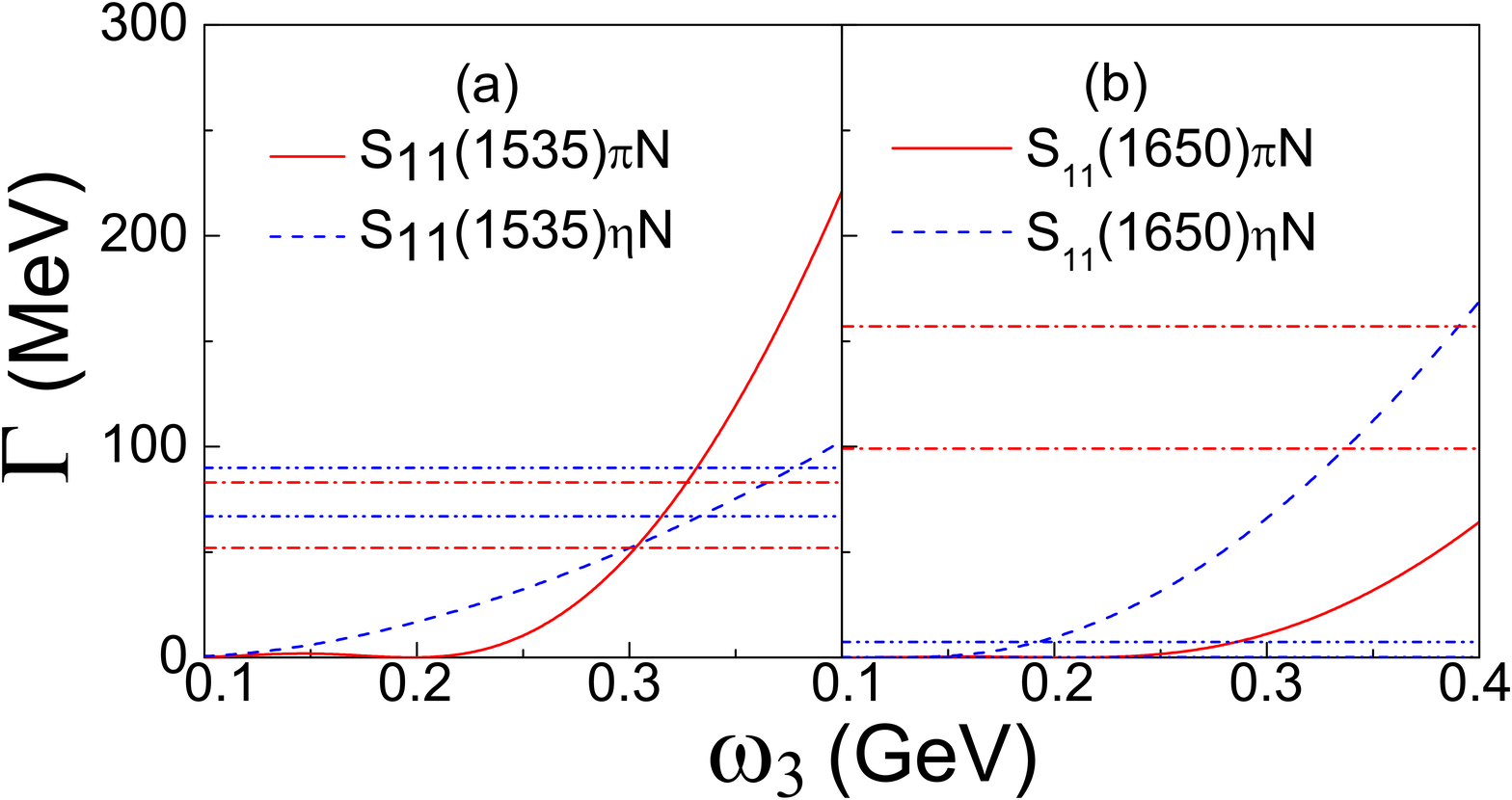}
\end{center}
\caption{(Color online) \footnotesize Partial decay widths of $S_{11}(1535)$ and
$S_{11}(1650)$ to $\pi N$ and $\eta N$ channels as a function of 
the harmonic oscillator parameter $\omega_{3}$.
Results of the present work are depicted in full and dashed curves for 
$S_{11} \to \pi N$ and $S_{11} \to \eta N$, respectively, without the
$SU(6)\otimes O(3)$ breakdown effects. The horizontal lines are the bands
given in PDG, for $S_{11} \to \pi N$ (dash-dotted) and $S_{11} \to \eta N$
(dash-dot-dotted).
\label{omega}}
\end{figure}
%
Figure~\ref{omega} shows the decay widths for $S_{11}(1535) \to \pi N,~\eta N$ (left panel)
and $S_{11}(1650) \to \pi N,~\eta N$ (right panel) as a function of $\omega_{3}$. 
The full and dashed curves are our results and the horizontal lines give the bands reported 
in PDG~\cite{Nakamura:2010zzi}.

The width for $S_{11}(1535) \to \pi N$ (full curve) falls in the experimental range (dash-dotted lines)
for $ 300 \lesssim \omega_{3} \lesssim 340$ MeV, while for $S_{11}(1535) \to \eta N$ the dashed curve and 
dash-dot-dotted lines lead to $ 300 \lesssim \omega_{3} \lesssim 380$ MeV. 
Accordingly, in the former range for $\omega_{3}$, the simple $qqq$ configuration allows reproducing 
the decay widths of $S_{11}(1535)$ in both $\pi N$ and $\eta N$ channels.

The situation with respect to the second $S_{11}$ resonance is dramatically different.
In the whole $\omega_{3}$ range, the calculated $S_{11}(1650) \to \pi N$ width (full curve), underestimates
the experimental band (dash-dotted lines). For the $\eta N$ decay channel, predicted values (dashed curve) 
agree with experimental band (dash-dot-dotted lines) below $\omega_{3} \approx 200$ MeV, where  
$\Gamma_{S_{11}(1650) \to \pi N}$ turns out vanishing. 

It is also worthwhile mentioning that, within exact $SU(6)\otimes O(3)$ symmetry, 
$\Gamma_{S_{11}(1650) \to K \Lambda}=0$ and hence, disagrees with the experimental value~\cite{Nakamura:2010zzi}: 
$4.8 \pm 0.7$ MeV.

In summary the pure $qqq$ configuration, within exact $SU(6)\otimes O(3)$, is not appropriate in describing
the  $S_{11}(1650)$ resonance properties. Consequently, one has to consider the
$SU(6)\otimes O(3)$ breakdown effects.

\subsection{Pure $qqq$ configuration and broken $SU(6)\otimes O(3)$ symmetry}
\label{num3qb}

As discussed in section \ref{wfm}, $SU(6)\otimes O(3)$ symmetry breaking effects
can be related to the mixing angles $\theta_{S}$ and $\theta_{D}$.
Several predictions on those angles are available (for a recent review see e.g. Ref.~\cite{Saghai:2009zz}).
Here, we will extract ranges for both angles and discuss them with respect to the two most
common approaches  leading to $SU(6)\otimes O(3)$ symmetry breaking, namely,
one-gluon-exchange (OGE)~\cite{Isgur:1977ef,Isgur:1978xb,Isgur:1978xi,Isgur:1978xj,Isgur:1978wd,Koniuk:1979vy}  
and one-boson-exchange models (OBE)~\cite{Glozman:1995fu}. 
Those approaches have raised some controversy~\cite{Isgur:1999jv,Glozman:1999ms}. 
Given that both the sign and the magnitude of the mixing angles in those approaches are different (see e.g. Refs.~\cite{Saghai:2009zz,Chizma:2002qi}), and that even within a given approach,
the sign depends on the convention used ~\cite{Saghai:2009zz,Koniuk:1979vy} or on the exchanged mesons
included~\cite{Glozman:1998wk}, we give in Appendix~\ref{apdx:mix}
values obtained within each approach in line with the de Swart~\cite{de Swart:1963gc} convention for SU(3).

In order to investigate the sign and range for $\theta_{S}$, 
in this section we report our numerical results for partial decay widths of $S_{11}(1535)$ and 
$S_{11}(1650)$ to ${\pi N}$ and $\eta N$ as a function of $\omega_3$ for six values of $\theta_{S}$, 
namely, $\pm 15^\circ,~\pm 30^\circ,~\pm 45^\circ$, and compare them to the data ranges.

\begin{figure}[ht]
\begin{center}
\includegraphics[scale=0.20]{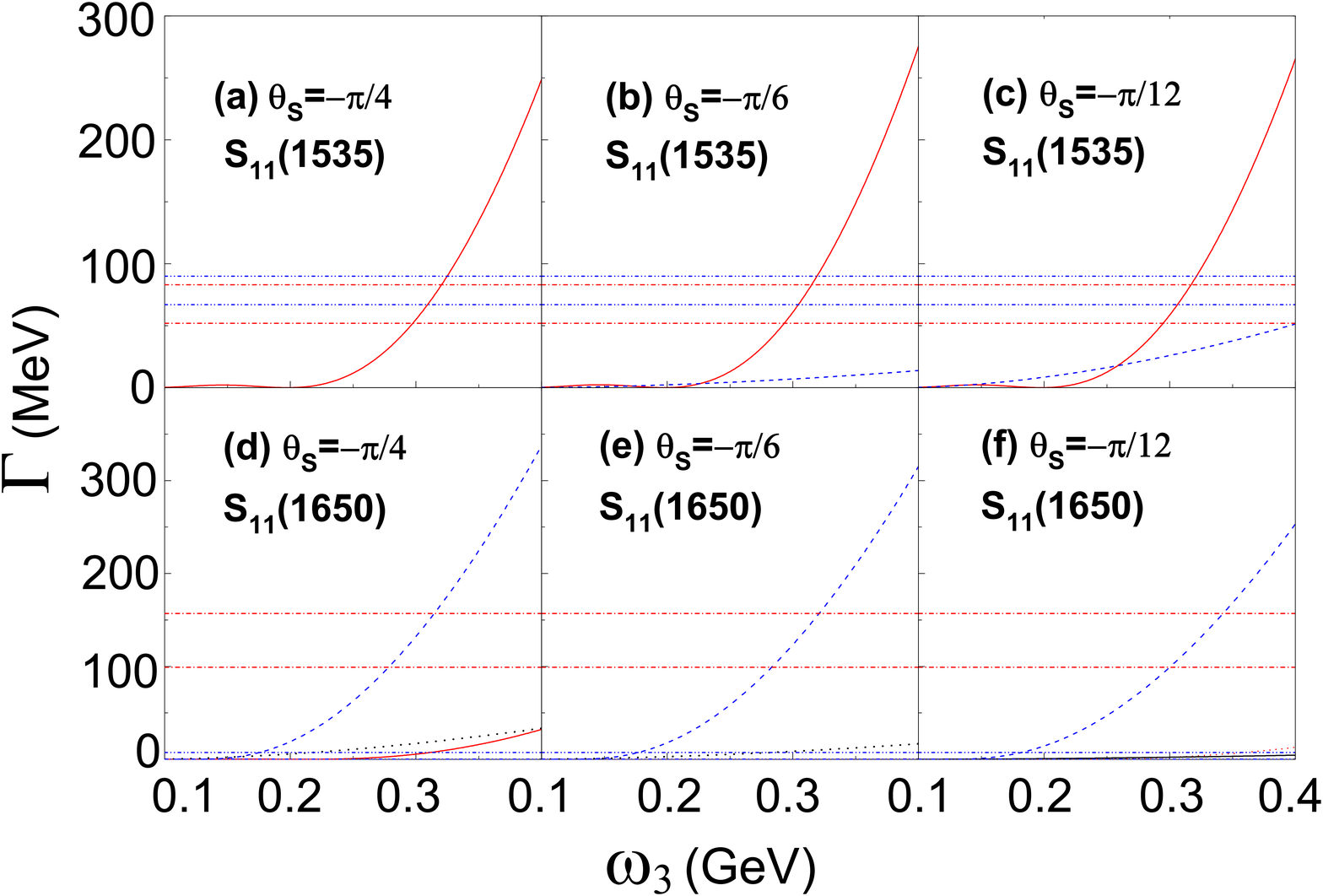}
\end{center}
\vspace{-1cm} 
\caption{(Color online) \footnotesize Decay widths of $S_{11}(1535)$ (upper
panel) and $S_{11}(1650)$ (lower panel) as a function of
$\omega_{3}$, with $\theta_{S}$ taken to be $- \frac{\pi}{4}$,
$- \frac{\pi}{6}$ and $ - \frac{\pi}{12}$, respectively. 
The dotted curves are our results for $\Gamma_{S_{11}(1650) \to K\Lambda}$, and the other ones are as 
in Fig.~\ref{omega}.
\label{minus}}
\end{figure}

In Fig.~\ref{minus} the strong decay partial widths $\Gamma_{S_{11} \to \pi N}$ and 
$\Gamma_{S_{11} \to \eta N}$  for $S_{11}(1535)$ (upper panel) and $S_{11}(1650)$ (lower panel) 
are shown as a function of $\omega_{3}$, with negative values for $\theta_{S}$. 
Conventions for the curves are the same as in Fig.~\ref{omega}, and due to 
$SU(6)\otimes O(3)$ symmetry breaking, $\Gamma_{S_{11}(1650) \to K\Lambda}$ gets non-vanishing values, depicted in
dotted curves. The experimental bands for this latter width are not shown, because they are almost identical to
those for $\Gamma_{S_{11}(1650) \to \eta N}$.

At all the three mixing angles, our predictions for $\Gamma_{S_{11}(1535) \to \pi N}$ and 
$\Gamma_{S_{11}(1650) \to \eta N}$ fall in the experimental bands for $\omega_3 \approx$ 300 MeV,
while the model underestimates very badly $\Gamma_{S_{11}(1535) \to \eta N}$ and 
$\Gamma_{S_{11}(1650) \to \pi N}$. 
Accordingly, within our approach, negative values for $\theta_{S}$ lead to unacceptable results 
compared to the data. 

%
%
\begin{figure}[hb!]
\begin{center}
\includegraphics[scale=0.20]{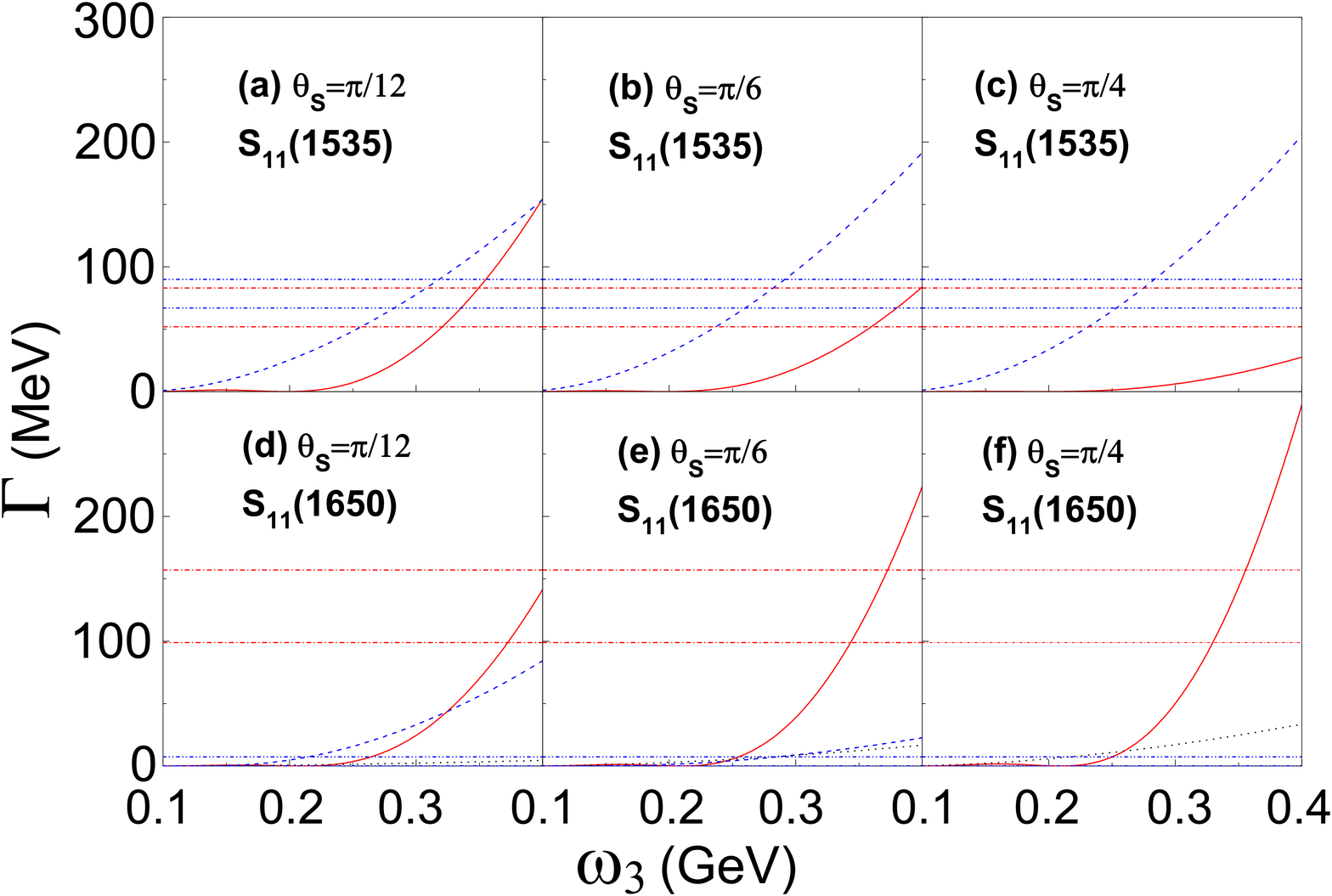}
\end{center}
\vspace{-0.5cm} 
\caption{(Color online) \footnotesize Same as Fig.~\ref{minus}, but with $\theta_{S}$ taken to be 
$\frac{\pi}{12}$, $\frac{\pi}{6}$ and $\frac{\pi}{4}$.  
\label{posit}}
\end{figure}

In Fig.~\ref{posit} the strong decay partial widths $\Gamma_{S_{11} \to \pi N}$ and $\Gamma_{S_{11} \to \eta N}$
for both $S_{11}$ resonances, as well as $\Gamma_{S_{11}(1650) \to K\Lambda}$, are depicted as a function 
of $\omega_{3}$ with positive values for $\theta_{S}$.
For the $S_{11}(1535)$ resonance, we obtain good agreement with data for $\theta_{S}= 15^\circ$ and 
$\omega_3 \approx 300$ MeV, for both ${\pi N}$ and ${\eta N}$ decay widths. 
This is also the case at all angles for $S_{11}(1650) \to {\pi N}$, but for $\omega_3 \approx 350$ MeV.

To go further in our investigation, we fix the harmonic  parameter at 
$\omega_3 = 340$ MeV and calculate partial widths and coupling constants for two extreme positive values
of the mixing angle, $\theta_{S}= 15^\circ$ and $35^\circ$. Moreover, we extend our study to
the $D_{13}(1520)$ and $D_{13}(1700)$ resonances, with the relevant mixing angle, also at two extreme values,
$\theta_{D}= 0^\circ$ and $17.5^\circ$. Results obtained within this procedure are hereafter
referred to as model A.

\begin{table}[ht]
\caption{\footnotesize Strong decay partial widths (in MeV) for the $S_{11}$ and $D_{13}$ resonances
in the three-quark model, with broken $SU(6)\otimes O(3)$ symmetry.}
\begin{tabular}{lccccccccccc}
\hline \hline
$N^*$          & $\Gamma_{tot}$ && $\pi N$ && $\eta N$   && $K \Lambda$   && Ref.   \\
\hline
$S_{11}(1535)$ & 150 $\pm$ 25 && 68 $\pm$ 15 && 79 $\pm 11$     &&  &&  PDG~\cite{Nakamura:2010zzi}  \\
               &           && 51 $\pm$ 21 &&121 $\pm$ 15 &&  && Model A \\ \\
$S_{11}(1650)$ & 165 $\pm$ 20 && 128 $\pm$ 29 && 3.8 $\pm$ 3.6  && 4.8 $\pm$  0.7 && PDG~\cite{Nakamura:2010zzi} \\
               &               && 81 $\pm$ 22 && 28 $\pm$ 22     && 9 $\pm$ 6    &&  Model A \\ \\
$D_{13}(1520)$ & 115 $\pm$ 15 && 69 $\pm$ 6  && 0.26 $\pm$ 0.05  &&           &&  PDG~\cite{Nakamura:2010zzi}  \\
               &           && 66 $\pm$ 7  && 0.19 $\pm$ 0.01&&           && Model A \\  
               &           && 72 $\pm$ 11  && 0.26 $\pm$ 0.07&&           && Jayalath {\it et al.}~\cite{Jayalath:2011uc} \\  \\
$D_{13}(1700)$ & 100 $\pm$ 50 &&  10 $\pm$ 5   &&  0.5 $\pm$  0.5   && 1.5  $\pm 1.5$   && PDG~\cite{Nakamura:2010zzi} \\
               &          && 13 $\pm$ 10    &&  0.5 $\pm$ 0.5 && 0.1 $\pm$ 0.1   &&  Model A \\
               &          && 12 $\pm$ 13    &&  $\leq$ 0.15  && $\leq$ 0.03 &&  Jayalath {\it et al.}~\cite{Jayalath:2011uc} \\
\hline \hline
\end{tabular}
\label{BRD13}
\end{table}

In Table~\ref{BRD13}, we present our results for the strong decay partial widths $\Gamma_{\pi N}$, 
$\Gamma_{\eta N}$ and $\Gamma_{K \Lambda}$ for the low lying $S_{11}$ and $D_{13}$ resonances studied here.

Within model A, the reduced $\chi^2$ per data point is 10.3. However, this large value is due to 
$S_{11} \to \eta N,~ K \Lambda$ decay channels. 
It is worthwhile mentioning that for the five $D_{13}$ partial decay widths, we get $\chi^2_{d.p.}$ = 0.7.

Here, $\Gamma_{S_{11}(1535) \to \pi N}$ is well reproduced, while $\Gamma_{S_{11}(1535) \to \eta N}$ is overestimated 
at the level of 3$\sigma$ and $\Gamma_{S_{11}(1650) \to \pi N}$ underestimated by roughly 2$\sigma$. 
For the remaining two other channels, large uncertainties on $\Gamma_{S_{11}(1650) \to \eta N}$ 
(both experiment and model),
and on $\Gamma_{S_{11}(1650) \to K \Lambda}$ (mainly model) do not lead to reliable conclusions.
Because of those undesirable features, we postpone to the next section the discussion on results from other sources, 
as well as the extraction of coupling constants.

For both $D_{13}$ resonances, the model A allows reproducing satisfactorily enough (Table~\ref{BRD13}) 
the known partial widths,
and agrees with values obtained within the $1/N_C$ expansion framework~\cite{Jayalath:2011uc}. 
model A is hence appropriate to put forward predictions for $D_{13}$-meson-baryon coupling constants.
In Table~\ref{ccd13}, our predictions for $\Gamma_{D_{13}MB}$ for seven meson-baryon sets 
are reported.
%
{\squeezetable
\begin{table}[ht]
\caption{\footnotesize Coupling constants for $D_{13}$ resonances to pseudoscalar meson and octet baryon within model A.}

\begin{tabular}{lccccccccc}
\hline \hline
$N^*$           && $\pi^{0}p$ & $\pi^{+}n$ & $\eta p$   & $K^{+}\Lambda$  & $K^{0}\Sigma^{+}$ & $K^{+}\Sigma^{0}$  & $\eta^{\prime}p$ \\
\hline
$D_{13}(1520)$  && -1.51 $\pm$ 0.07 & 2.13 $\pm$ 0.10  &  -8.33 $\pm$ 0.20 & 3.44 $\pm$ 0.08
               & 0.99 $\pm$ 0.14 & -0.69 $\pm$ 0.09 & 2.11 $\pm$ 0.05  \\
$D_{13}(1700)$ && -0.35 $\pm$ 0.17     &  0.50 $\pm$ 0.25 & 0.93 $\pm$ 0.91& 1.43 $\pm$ 1.43
               & -2.80 $\pm$ 0.05     & 1.98 $\pm$ 0.04 &  1.67 $\pm$ 0.52 \\
\hline \hline
\end{tabular}
\label{ccd13}
\end{table}
}

To end this section, we summarize our main findings within a traditional $qqq$
$\chi$CQM, complemented with $SU(6)\otimes O(3)$ breakdown effects, and using following input values
for adjustable parameters: $\omega _3$ = 340 MeV, $15^\circ \leq \theta_{S} \leq 35^\circ$ and 
$0^\circ\leq \theta_{D}\leq 17.5^\circ$.

Model A is found appropriate for the $D_{13}$ resonances, given that the partial decay widths show from reasonable 
to good agreements with the PDG values.
So, we do not push further our studies with respect to the $D_{13}(1520)$ and  $D_{13}(1700)$.

The main shortcomings of the model A concern: $\Gamma_{S_{11}(1535) \to \eta N}$ and the fact that for the $S_{11}(1650)$ 
resonance, central values for all three channels show significant discrepancies with those reported in PDG.
This latter point remains problematic because of large uncertainties. 

Attempting to cure those disagreements with respect to the $S_{11}$ resonances, we proceed in the next section 
to considering possible contributions from higher Fock-components.


\subsection{Mixed $qqq$ and $qqqq \bar q$ configuration and broken $SU(6)\otimes O(3)$ symmetry}
\label{numtt}
To produce numerical results, seven input parameters are needed, the values of which are discussed below.

{\it a) Constituent quarks' masses:} 
due to the introduction of five-quark components, masses to be used are smaller than those we adopted in section \ref{num3q}, 
while dealing with pure three-quark states.
In line with Ref.~\cite{An:2008xk}, we take $m=290$ MeV and $m_{s}=430$ MeV. 

{\it b) Oscillator parameters:}
following results presented in section~\ref{num3q}, we fix the oscillator parameter at $\omega_{3}=340$ MeV.
For the five-quark components a commonly used value for the oscillator parameter, $\omega_{5}=600$ MeV,
is adopted.

{\it c) Mixing angle:}
in Section~\ref{num3qb}, we showed that to fit the decay widths , the mixing angle should be in the range
$15^\circ \leq \theta_{S} \leq 35^\circ$. 
In the following, this angle is treated as adjustable parameter.

{\it d) Probabilities of five-quark components:} 
the probabilities of the five-quark components in $S_{11}(1535)$ ($P_{5q}=A_{5q}^{2}$) and $S_{11}(1650)$
($P^{\prime}_{5q}=A^{\prime ^2}_{5q}$) are also adjustable parameters in our model search.

The latter three adjustable parameters have been extracted by mapping out the whole phase space
defined by $15^\circ \leq \theta_{S} \leq 35^\circ$ and from 0 to 100\% for five-quark probabilities
in both $S_{11}(1535)$ and $S_{11}(1650)$. The calculated observables are: the partial decay widths 
of both $S_{11}$ resonances to $\pi N$ and $\eta N$, as well as $\Gamma_{S_{11}(1650) \to K \Lambda}$.
Sets [$\theta_{S},~P_{5q},~P^{\prime}_{5q}$] leading~\cite{An:2011nu} to decay widths within ranges 
reported in PDG have been singled out. 
Then, for each partial widths, extreme values for those parameters are retained as model ranges, namely,
\begin{equation}
26.8^\circ \leq \theta_{S} \leq 29.8^\circ~;~
21\% \leq P_{5q} \leq 30\%~;~
11\% \leq P^\prime_{5q} \leq 18\%\,.
\label{Ranges}
\end{equation}
The obtained model is hereafter called model B.

\begin{figure}[t]
\begin{center}
\includegraphics[scale=0.23]{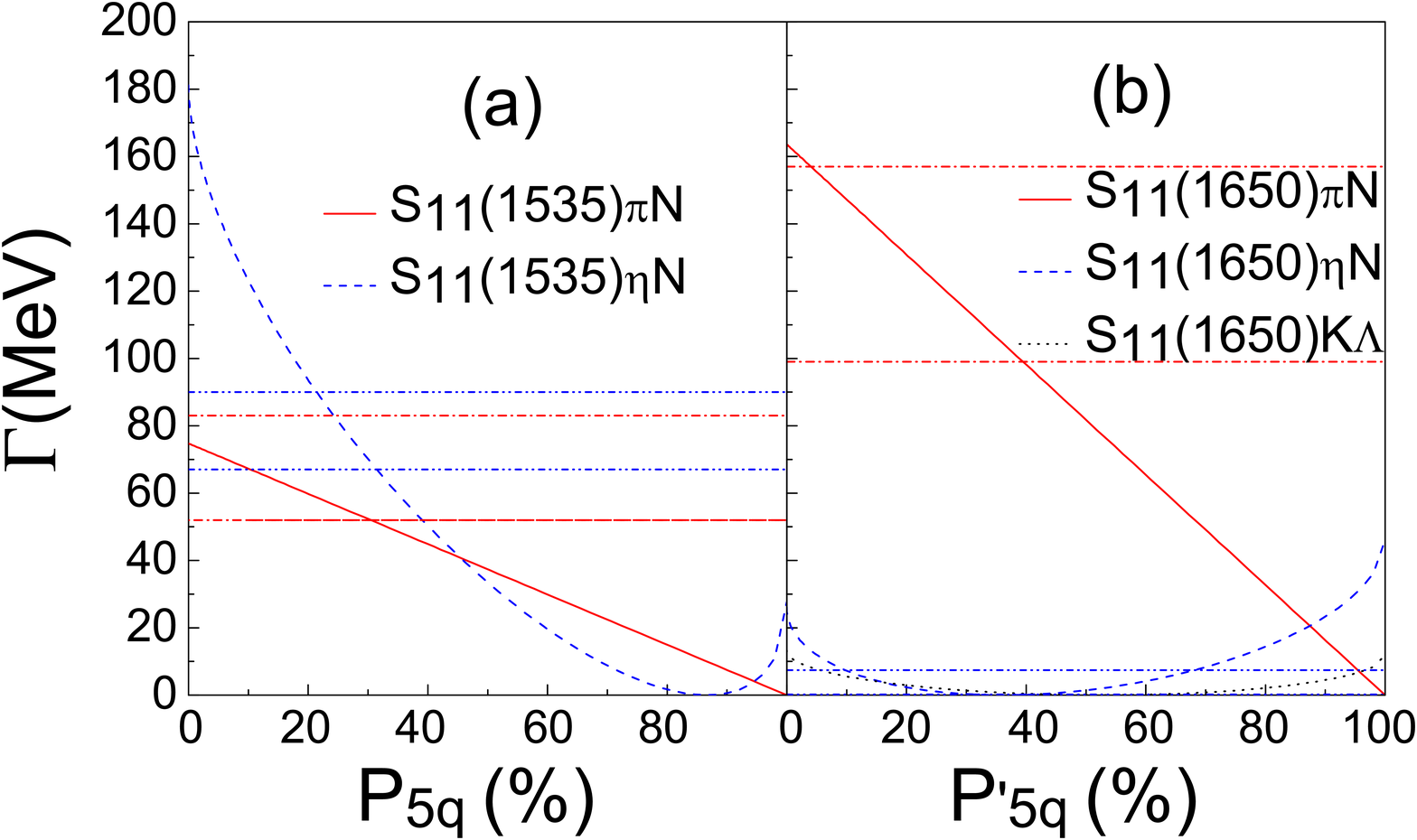}
\end{center}
\vskip -1.0 cm
\caption{(Color online) \footnotesize Partial decay widths (in MeV) for $S_{11}$ resonances as a function of five-quark
components, $\theta _S=28^\circ$. Curves the same as in Fig.~\ref{posit}.
\label{prob}}
\end{figure}

As an example, Fig.~\ref{prob} illustrates how the known ranges for the partial decay widths allow determining 
ranges for the five-quark components' probabilities. 
There, for each decay width intersections of the model curve with the horizontal bands taken from PDG, determine 
 the extreme values for the relevant five-quark probability.
 
%
 Notice that the probability range for five-quark component in $S_{11}(1535)$ given above is compatible
 with previous results~\cite{An:2008xk,An:2008tz}, obtained within $\chi CQM$ approaches.
 The latter one~\cite{An:2008tz} puts an upper limit of $P_{5q} \leq 45\%$, 
 based on the axial charge study of the resonance.
 While the former one~\cite{An:2008xk}, dedicated to the electromagnetic transition $\gamma ^* N \to S_{11}(1535)$, 
 reports $25\% \leq P_{5q} \leq 65\%$. 
\subsubsection{Partial decay widths $\Gamma_{S_{11} \to MB}$} 
The resulting numerical partial decay widths, within both models A and B, are reported in Table~\ref{BR} and compared
with the PDG data~\cite{Nakamura:2010zzi} as well as with results from other authors, based on various 
approaches~\cite{Penner:2002ma,Inoue:2001ip,Aznauryan:2003zg,Vrana:1999nt,Shyam:2007iz,Arndt:2005dg,Ceci:2006jj,Golli:2011jk,Jayalath:2011uc}.

Comparing results of the models A and B with the data for all five channels, shows clearly the superiority of
the model B. The $\chi^2_{d.p.}$ is 0.15, instead of 19.9 in the case of model A.

The most striking feature here is that $\Gamma_{S_{11}(1535) \to \eta N}$ is nicely reproduced, which was not the case
with previous configurations, namely, pure $qqq$ without or with $SU(6) \otimes O(3)$ symmetry breaking. 
Moreover, $\Gamma_{S_{11}(1535) \to \pi N}$ agrees with PDG values within better than 1$\sigma$.  
The range for $\Gamma_{S_{11}(1650) \to \pi N}$ gets significantly reduced within the model B with respect to the 
model A result and is compatible with the PDG value within less than 1$\sigma$. 
Narrow experimental widths for  
$\Gamma_{S_{11}(1650) \to \eta N}$ and $\Gamma_{S_{11}(1650) \to K\Lambda}$ are well reproduced by the model B,
with uncertainties comparable to those of the data. In the following, we proceed to comparisons with
results from other sources.

The most complete set of results comes from a very recent comprehensive study~\cite{Jayalath:2011uc} of all
known partial decay widths for sixteen baryon resonances, within the framework of the $1/N_C$ expansion
in the next to leading order (NLO) approximation. 
Results for the $S_{11}(1535)$ decay channels from that work and model B are in excellent agreement.
For the $S_{11}(1650)$, given that the authors of Ref.~\cite{Jayalath:2011uc} use branching fractions data
in PDG for $\eta N$ and $K \Lambda$ channels, rather than the branching ratios, we postpone the
comparisons to sec.~\ref{brbf}.
 
%
{\squeezetable
\begin{table}[t]
\caption{\footnotesize Strong decay widths (in MeV) for $S_{11}(1535)$ and $S_{11}(1650)$.}
\begin{tabular}{lccccccccccc}
\hline \hline
$N^*$          & $\Gamma_{tot}$ && $\pi N$ && $\eta N$&& $K \Lambda$ && Approach && Ref.   \\
\hline
$S_{11}(1535)$ & 150 $\pm$ 25 && 68 $\pm$ 15 && 79 $\pm$ 11 &&   &&                 && PDG~\cite{Nakamura:2010zzi}  \\
               &              && 51 $\pm$ 21  &&121 $\pm$ 15 &&   && Model A         && Present work \\
               &              && 58 $\pm$ 5 && 79 $\pm$ 11 &&       && Model B         &&  Present work \\
               &              && 57 $\pm$19 && 73 $\pm$ 44  &&    && $1/N_C$-NLO && Jayalath~{\it et al.}~\cite{Jayalath:2011uc} \\
               &112 $\pm$ 19&& 39 $\pm$ 5 && 57 $\pm$ 6  &&    && Coupled-channel && Vrana~{\it et al.}~\cite{Vrana:1999nt} \\
               &129 $\pm$ 8&& 46 $\pm$ 1 && 68 $\pm$ 1  &&    && Coupled-channel && Penner-Mosel~\cite{Penner:2002ma} \\
               &136        && 34.4       && 56.2        &&    && Coupled-channel && Shyam~\cite{Shyam:2007iz} \\
               &           && 42 $\pm$ 6 && 70 $\pm$ 10 &&       &&  PWA            && Arndt~{\it et al.}~\cite{Arndt:2005dg}  \\
               &           && 21.3       && 65.7     &&        &&  Chiral Unitary && Inoue~{\it et al.}~\cite{Inoue:2001ip}  \\
               &  95       && 42         && 51          &&    && Chiral quark model && Golli~{\it et al.}~\cite{Golli:2011jk} \\
               & 165       && 64         && 89          &&    &&  K-Matrix       && Ceci~{\it et al.}~\cite{Ceci:2006jj}  \\
               & 142       &&            && 71       &&        &&  Disp. Rel.     && Aznauryan~\cite{Aznauryan:2003zg}  \\
               & 195       &&            && 97       &&        &&  Isobar         && Aznauryan~\cite{Aznauryan:2003zg}  \\ \\
$S_{11}(1650)$ & 165 $\pm$ 20 &&128 $\pm$ 29&& 3.8 $\pm$ 3.6 && 4.8 $\pm$  0.7 &&          && PDG~\cite{Nakamura:2010zzi}  \\
               &              && 81 $\pm$ 22&& 28 $\pm$ 22   && 9 $\pm$ 6      &&  Model A && Present work \\
               &              &&143 $\pm$ 5 && 4.5 $\pm$ 3.0 && 4.8 $\pm$ 0.7  &&  Model B &&  Present work \\
               &202 $\pm$ 40&& 149 $\pm$ 4 && 12 $\pm$ 2  &&    && Coupled-channel && Vrana~{\it et al.}~\cite{Vrana:1999nt} \\
               & 138 $\pm$ 7  && 90 $\pm$ 6 && 1.4 $\pm$ 0.8 && 3.7 $\pm$ 0.6  &&  Coupled-channel && Penner-Mosel~\cite{Penner:2002ma} \\
               &133          && 71.9        && 2.5           &&    && Coupled-channel && Shyam~\cite{Shyam:2007iz} \\
               &   	144       && 86         && 1.4           &&  13            &&  Chiral quark model && Golli~{\it et al.}~\cite{Golli:2011jk} \\
               & 233          && 149        && 37            &&                &&  K-Matrix        &&  Ceci~{\it et al.}~\cite{Ceci:2006jj}  \\
               &  85          &&            && 3.2           &&                &&  Disp. Rel.      &&  Aznauryan~\cite{Aznauryan:2003zg}  \\
               & 125          &&            && 6.9           &&                &&  Isobar          &&  Aznauryan~\cite{Aznauryan:2003zg}  \\
\hline \hline
\end{tabular}
\label{BR}
\end{table}
}
The Pitt-ANL~\cite{Vrana:1999nt} multichannel analysis of $\pi N \to \pi N,~\eta N$, 
produces rather small total widths for
$S_{11}(1535)$ and large one for $S_{11}(1650)$. Those features lead to underestimate of 
$\Gamma_{S_{11}(1535) \to \pi N}$ and $\Gamma_{S_{11}(1535) \to \eta N}$, and overestimate of 
$\Gamma_{S_{11}(1650) \to \eta N}$. However, $\Gamma_{S_{11}(650) \to \pi N}$ comes out in agreement with
PDG and model B results.

An extensive coupled-channels analysis~\cite{Penner:2002ma,Penner:2002md} studied within an
isobar approach all available data by year 2002 for following processes:
$\gamma N \to \gamma N$, $\pi N$, $\pi \pi N$, $\eta N$,~$K \Lambda$, $K \Sigma$, $\omega N$
and $\pi N \to \pi N$, $\eta N$, $K \Lambda$, $K \Sigma$, $\omega N$.
That work describes successfully four out of the five decay channels, albeit with a few tens of free parameters, 
with the main shortcoming being the underestimate of $\Gamma_{S_{11}(1535) \to \pi N}$.

Interpreting $p N \to p N \eta$ data, within an effective Lagrangian approach~\cite{Shyam:2007iz}, 
underestimates all partial decay widths, except $\Gamma_{S_{11}(1650) \to K \Lambda}$.

The latest available results from SAID~\cite{Arndt:2005dg}, in 2005, analyzing $\pi N$ elastic scattering and 
$\eta N$ production data, give a smaller $\Gamma_{S_{11}(1535) \to \pi N}$ with respect to PDG, and compatible 
with PDG value for $\Gamma_{S_{11}(1535) \to \eta N}$. 

A chiral unitary approach~\cite{Inoue:2001ip} dedicated to the $S$-wave meson-baryon interactions, 
reproduces well $\Gamma_{S_{11}(1535) \to \eta N}$, but underestimates $\Gamma_{S_{11}(1535) \to \pi N}$ by more 
than a factor of 2.

A recent chiral quark model~\cite{Golli:2011jk}, concentrating on the meson scattering and 
$\pi$ and $\eta$ electroproduction amplitudes,
leads to rather small total width for both resonances, underestimating all $\pi N$ and $\eta N$ partial decay widths 
by roughly 2$\sigma$, and overestimating $\Gamma_{S_{11}(1650) \to K \Lambda}$ by more than 10$\sigma$.
The authors conclude however that the $S_{11}(1535)$ resonance is dominated by a genuine three-quark state.

Results of a K-matrix approach~\cite{Ceci:2006jj} for $\pi N$ and $\eta N$ final states provide realistic values
for all considered partial widths, except for $\Gamma_{S_{11}(1650) \to \eta N}$.

Finally, in Ref.~\cite{Aznauryan:2003zg}, studying the $\eta N$ final states, dispersion relations lead to values 
in agreement with data, while the isobar model tends to overestimate $\Gamma_{S_{11}(1535) \to \eta N}$.

The ambitious EBAC~\cite{EBAC} program offers a powerful frame to study the properties of baryons, including 
partial decay widths~\cite{Kamano:2008gr}, extraction of which requires non ambiguous determination of the 
poles positions~\cite{Suzuki:2008rp}; a topic under extensive 
investigations~\cite{Suzuki:2008rp,Capstick:2007tv,Doring:2009bi,Doring:2009yv,Doring:2009uc,Oset:2009vf,Ceci:2011ae,Osmanovic:2011xn}.
\subsubsection{Coupling constants $g_{S_{11}MB}$}
\label{2cc}
In Table~\ref{cc}, predictions for the relevant resonance-meson-baryon coupling constants, $g_{S_{11}MB}$, 
from models A and B are given in particle basis. 

In order to emphasize the most sensitive decay channels to the five-quark components in $S_{11}(1535)$, we compare
results from models A and B. For $K^{+}\Sigma^{0}$ and $K^{0}\Sigma^{+}$, we observe variations by a factor
of 2 between the two models, with central values differing from each other by more than 4$\sigma$.
Next come $K^{+}\Lambda$ and $\eta p$, with about 30\% differences and 2$\sigma$. 
The other three channels ($\pi^{0}p$, $\pi^{+}n$, $\eta^{\prime}p$) show no significant sensitivities to the
five-quark components.

In the case of $S_{11}(1650)$, similar sensitivities are observed. However, the
rather small branching ratios to those final states, require substantial experimental efforts and sophisticated
phenomenological approaches, e.g. for $\gamma p \to K^{0}\Sigma^{+},~ K^{+}\Sigma^{0}$.
  
In Table~\ref{cc}, results from a chiral unitary approach~\cite{Inoue:2001ip} are also reported, showing 
compatible values with those of model B for $K^{+}\Sigma^{0}$, $K^{0}\Sigma^{+}$ and $\eta p$. For the 
other three channels the two sets differ by roughly 60\%.
%
{\squeezetable
\begin{table}[t]
\caption{\footnotesize $S_{11}$-meson-baryon coupling constants ($g_{S_{11} MB}$) in particle basis.}

\begin{tabular}{lccccccccc}
\hline \hline
$N^*$           && $\pi^{0}p$ & $\pi^{+}n$ & $\eta p$   & $K^{+}\Lambda$  & $K^{0}\Sigma^{+}$ & $K^{+}\Sigma^{0}$  & $\eta^{\prime}p$  & Ref.   \\
\hline
$S_{11}(1535)$               && -0.58 $\pm$ 0.13 & 0.82 $\pm$ 0.18  & -2.57 $\pm$ 0.17     &   1.42 $\pm$ 0.11
               & 0.95 $\pm$ 0.20 &-0.62 $\pm$ 0.09 & 3.09 $\pm$ 0.20                & Model A\\ 
               && -0.63 $\pm$ 0.03 & 0.89 $\pm$ 0.04  &  -2.07 $\pm$ 0.15     & 1.76 $\pm$ 0.02
               & 1.81 $\pm$ 0.06 & -1.28 $\pm$ 0.04 & 3.33 $\pm$ 0.10               & Model B\\
               && $\pm0.39$ & $\pm0.56$  &  $\pm1.84$& $\pm0.92$& $\pm2.12$ &$\pm1.50$ && \cite{Inoue:2001ip}\\ \\
$S_{11}(1650)$                && -0.70 $\pm$ 0.10     &  0.94 $\pm$ 0.19 & 0.84 $\pm$ 0.40 &  0.67 $\pm$ 0.25
               & -1.42 $\pm$ 0.21     & 0.95 $\pm$ 0.10&  -1.61 $\pm$ 0.79      &Model A\\

               && -0.94 $\pm$ 0.02     & 1.33 $\pm$ 0.03 &  0.35  $\pm$ 0.12 & 0.51 $\pm$ 0.03
               & -2.17 $\pm$ 0.05    &1.53 $\pm$ 0.04 &  -1.62 $\pm$ 0.14      &Model B\\
\hline \hline
\end{tabular}
\label{cc}
\end{table}
}
%
{\squeezetable
\begin{table}[th!]
\caption{\footnotesize $S_{11}$-meson-baryon coupling constants ($g_{S_{11} MB}$) in isospin basis.}
\begin{tabular}{lcccccccccccccc}
\hline \hline
$N^*$          &&        $\pi N$   &&      $\eta N$    &&    $K \Lambda$   && $K \Sigma$ && $\eta^\prime N$ && Approach && Ref.   \\
\hline
$S_{11}(1535)$ && -1.09 $\pm$ 0.05 && -2.07 $\pm$ 0.15 &&  1.76 $\pm$ 0.02 && 2.21 $\pm$ 0.07 && 3.3 $\pm$ 0.1 && Model B && Present work \\
               &&$\pm$(0.62$\pm$0.32) &&$\pm$(0.97$\pm$0.45) && $\pm$(0.55$\pm$0.32) && $\pm$(0.55$\pm$0.32)&& && PWA &&  Sarantsev {\it et al.} \cite{Sarantsev:2005tg}\\
               && $\pm$0.6 && $\pm$2.1 &&  $\pm$1.7 &&  $\pm$2.4 && && Chiral Lagrangian &&  Gamermann {\it et al.} \cite{Gamermann:2011mq}\\\\
$S_{11}(1650)$ && -1.64 $\pm$ 0.03 && 0.35 $\pm$ 0.14 &&  0.53 $\pm$ 0.04 && -2.66 $\pm$ 0.06 && -1.62 $\pm$ 0.14 &&  Model B &&  Present work \\
               &&$\pm$(1.05$\pm$0.45) &&$\pm$(0.63$\pm$0.32) && $\pm$(0.32$\pm$0.32) && $\pm$(0.71$\pm$0.39)&& && PWA && Sarantsev {\it et al.} \cite{Sarantsev:2005tg}\\
               && $\pm$1.2 && $\pm$0.8 &&  $\pm$0.6 &&  $\pm$1.7 && && Chiral Lagrangian &&  Gamermann {\it et al.} \cite{Gamermann:2011mq}\\
\hline \hline
\end{tabular}
\label{cci}
\end{table}
}

In Table~\ref{cci}, predictions in isospin basis are reported for model B and other sources. 
Additional results reported in the literature and limited to fewer channels are also discussed below.

Within an isobar approach~\cite{Anisovich:2005tf}, a combined analysis~\cite{Sarantsev:2005tg} of the pseudoscalar
mesons photoproduction data available by 2005 has extracted coupling constants in isospin basis, 
with around $\pm$60\% uncertainties.
The reported couplings $g_{S_{11}(1535) \pi N}$ and $g_{S_{11}(1535) \eta N}$ are compatible with
the model B predictions within 2$\sigma$, while discrepancies between the two approaches for 
$g_{S_{11}(1535) K \Lambda}$ and $g_{S_{11}(1535) K \Sigma}$ reach factors 3 to 4 and 4$\sigma$.
For the second resonance, results from the two calculations agree within 1$\sigma$ for
$g_{S_{11}(1650) \pi N}$, $g_{S_{11}(1650) \eta N}$ and $g_{S_{11}(1650) K \Lambda}$, with only
significant disagreement observed for $g_{S_{11}(1650) K \Sigma}$.
Copious data released since then, if interpreted within the same approach might bring in new insights 
into the coupling constants.
  
Results from a recent SU(6) extended chiral Lagrangian~\cite{Gamermann:2011mq}, embodying eleven 
meson-baryon final states, are also reported in Table~\ref{cci} and show consistent values between that 
approach and model B for
$g_{S_{11}(1535) \eta N}$, $g_{S_{11}(1535) K \Lambda}$, $g_{S_{11}(1535) K \Sigma}$, and 
$g_{S_{11}(1650) K \Lambda}$.     
  
An effective Lagrangian focused on interpreting~\cite{Shyam:2007iz} $\eta$ production data in $NN$ and $\pi N$ 
collisions, leads to $g_{S_{11}(1535) \eta N}$ = 2.2 and $g_{S_{11}(1650) \eta N}$=0.55, compatible with our values.
Another effective Lagrangian approach~\cite{Cao:2008st} studying $\eta$ and $\eta^\prime$ production data in 
the same reactions gives  $g_{S_{11}(1535) \eta^{\prime}p}$ = 3.7, about only 10\% higher than the value 
given by model B.
        
Here, we wish to make a few comments with respect to the relative values of some of the coupling constants.

{\it i)} While the $\eta N N$ coupling constant is known to be smaller than that of $\pi N N$, 
the ratio $|g_{S_{11}(1535) \eta N}/g_{S_{11}(1535)\pi N}|$ comes out significantly larger than 1.
This result is in line with the finding~\cite{Jido:1997yk} that, in the soft pion limit, $\pi N N^*$ 
coupling vanishes due to chiral symmetry, while that of $\eta N N^*$ remains finite.

{\it ii)} The ratio $|g_{S_{11}(1535)K\Lambda}/g_{S_{11}(1535)\eta N}|$ takes the value $1.3\pm 0.3$, within an
isobar model~\cite{Liu:2005pm} interpreting $J/\psi\to \bar{p}p\eta$ and
$\psi \to \bar{p}K^{+}\Lambda$ data, larger than the results reported in Table~\ref{cci}.
Dressed versus bare mass considerations~\cite{Ceci:2009zz}, might affect the reported ratio 
in Ref.~\cite{Liu:2005pm}. 
Investigation of the same reaction within a unitary chiral approach~\cite{Inoue:2001ip,Geng:2008cv} puts
that ratio around 0.5 to 0.7, smaller than our result.

{\it iii)} The ratio $|g_{S_{11}(1650) K \Sigma}/g_{S_{11}(1650) K \Lambda}|$ turns out to be around 5.
 Actually, $S_{11}(1650)$ is dominant by the state $N(^{4}_{8}P_{M})_{\frac{1}{2}^{-}}$, which cannot transit 
 to $K\Lambda$ channel.  
 Moreover, there is a cancellation between the contributions from $qqq \to K \Lambda$ and 
 $qqqq \bar q \to K \Lambda$, which leads also to a very small decay width 
 $\Gamma_{S_{11}(1650) \to K \Lambda}$. 
In addition, the threshold for $S_{11}(1650) \to K\Sigma$ decay channel being very close to
the mass of $S_{11}(1650)$, contributions from the five-quark component enhance significantly the 
coupling constant $g_{S_{11}(1650)K\Sigma}$.

{\it iv)} It is worthy to be noticed that he coupling constants 
$g_{S_{11}\eta N}$, $g_{S_{11}K\Sigma}$ and $g_{S_{11}\eta^{\prime}N}$ for $S_{11}(1535)$ and 
$S_{11}(1650)$ have opposite signs. 
Moreover, the ratio $|g_{S_{11}(1535)K\Sigma}/g_{S_{11}(1650)K\Sigma}|$ is close to unity.
Those features might lead to significant cancellations in the interference terms in $KY$ photo- and/or 
hadron-induced productions.   
    
{\it v)} In Tables~\ref{cc} and ~\ref{cci}, one finds the following orderings for magnitudes of the coupling constants,
predicted by model B, and in Refs.\cite{Bruns:2010sv,Gamermann:2011mq}, noted below as {\bf a)}, 
{\bf b)} and {\bf c)}, respectively: 

{\bf - For} {\boldmath {$S_{11} \equiv S_{11}(1535)$:}}

$\bullet$ {\it In particle basis}
\begin{eqnarray}
{\bf (a)}&:&~|g_{S_{11}\pi^{0}p}| < |g_{S_{11}\pi^{+}n}| < |g_{S_{11}K^{+}\Sigma^{0}}| < |g_{S_{11}K^{+} \Lambda} |
\approx |g_{S_{11}K^{0}\Sigma^{+}}| < |g_{S_{11}\eta p}| < |g_{S_{11}\eta^{\prime}p}|, \\
{\bf (b)}&:&~|g_{S_{11}\pi^{0}p}| 
\approx |g_{S_{11}K^{+}\Sigma^{0}}| < |g_{S_{11}\pi^{+}n}| \approx |g_{S_{11}K^{0}\Sigma^{+}}|
< |g_{S_{11}\eta p}| < |g_{S_{11}K^{+} \Lambda} |.
\end{eqnarray} 

The main feature of our results {\bf (a)} is that the strongest couplings are found the hidden strangeness 
sector, while those for open strangeness channels come out in between $\pi N$ and $\eta N$ final states.

Inequalities in {\bf (b)} come from a recent unitarized chiral effective Lagrangian~\cite{Bruns:2010sv}, in which both 
$S_{11}(1535)$ and $S_{11}(1650)$ are dynamically generated. 
Within that model, the coupling to $K^{+}\Sigma^{0}$ is highly suppressed, and that to $K^{+} \Lambda$ turns out 
larger than coupling to $\eta p$.

$\bullet$ {\it In isospin basis}
\begin{eqnarray}
{\bf (a^\prime)}&:&~|g_{S_{11}\pi N}|  < |g_{S_{11}K \Lambda} | < |g_{S_{11}\eta N}| \approx |g_{S_{11}K \Sigma}| 
< |g_{S_{11}\eta^{\prime}N}|,\\
{\bf (c^\prime)}&:&~|g_{S_{11}\pi N}|  < |g_{S_{11}K \Lambda} | < |g_{S_{11}\eta N}| \approx |g_{S_{11}K \Sigma} |.
\end{eqnarray}
Results from a chiral Lagrangian study~\cite{Gamermann:2011mq}, {\bf (c')}, give the same ordering for couplings as 
model B. 
It is also the case for results from a chiral unitary approach~\cite{Inoue:2001ip}, 
while another chiral unitary approach~\cite{Hyodo:2008xr}, distinguishing dynamically generated resonances from 
genuine quark states, leads to
\begin{eqnarray}
|g_{S_{11}\pi N}|  < |g_{S_{11}K \Lambda} | < |g_{S_{11}\eta N}| < |g_{S_{11}K \Sigma}|.
\end{eqnarray}

{\bf - For} {\boldmath {$S_{11} \equiv S_{11}(1650)$:}}

$\bullet$ {\it In particle basis}
\begin{eqnarray}
{\bf (a)}&:&~|g_{S_{11}\eta p}| < |g_{S_{11}K^{+} \Lambda} | < |g_{S_{11}\pi^{0}p}| < |g_{S_{11}\pi^{+}n}| < 
|g_{S_{11}K^{+}\Sigma^{0}}| < |g_{S_{11}\eta^{\prime}p}| < |g_{S_{11}K^{0}\Sigma^{+}}|, \\  
{\bf (b)}&:&~|g_{S_{11}K^{+} \Lambda} | < |g_{S_{11}\pi^{0}p}| < |g_{S_{11}\pi^{+}n}| \approx |g_{S_{11}K^{+}\Sigma^{0}}|  
< |g_{S_{11}\eta p}| <  |g_{S_{11}K^{0}\Sigma^{+}}|.
\end{eqnarray}
In our model, the ordering in strangeness sector is separated by $\pi N$, according to the fact that the relevant 
disintegration channel is above or below the resonance mass.

The main differences between results from model B and those in Ref.~\cite{Bruns:2010sv} concern couplings to
$K^{+} \Lambda$ and $\eta p$.

$\bullet$ {\it In isospin basis}
\begin{eqnarray}
{\bf (a^\prime)}&:&~ |g_{S_{11}K \Lambda} | < |g_{S_{11}\eta N}| < |g_{S_{11}\pi N}|\approx |g_{S_{11}\eta^{\prime}N}| 
< |g_{S_{11}K \Sigma} |, \\  
{\bf (c^\prime)}&:&~ |g_{S_{11}}\eta N | \lesssim |g_{S_{11}K \Lambda}| < |g_{S_{11}\pi N}|  < |g_{S_{11}K \Sigma} |.
\end{eqnarray}
Here again model B and Ref.~\cite{Gamermann:2011mq} lead basically to identical orderings.

To end this section, we would like to emphasize the following point, with respect to the importance of
five-quark components.
Our model leads to probability for the strangeness component in $S_{11}(1650)$ being smaller than that for 
the five-quark component in $S_{11}(1535)$.
Moreover, the probability amplitude turns out to be positive for $S_{11}(1535)$, but negative for $S_{11}(1650)$.

Taking the ranges determined for probabilities (Eq.~(\ref{Ranges})), one gets $-77.4 \leq A_{5q}/A_{5q}^\prime \leq -72.5$. 
This latter range and that for $\theta_{S}$, embodied in Eq.~(\ref{ampratio}), allow extracting values for the 
energy of the strangeness component, $1641.60 \leq E_{5} \leq 1649.99$ MeV.
The coupling between $qqq$ and $qqqq\bar{q}$ in the corresponding baryon $_{5q}\langle\hat{V}_{cou} \rangle_{3q}$,
Eq.~(\ref{ampratio}), turns out to be negative for both $S_{11}$ resonances.
\subsubsection{Branching fraction versus branching ratio considerations}
\label{brbf}
As mentioned earlier, in PDG~\cite{Nakamura:2010zzi} estimates for both branching fractions (BF) to 
meson-baryon states and branching ratios (BR), ($\Gamma_{MB}/\Gamma_{total}$), are reported. 
In the case of the $S_{11}$ resonances considered here, those estimates are not identical for 
$S_{11}(1650) \to \eta N,~ K \Lambda$.
In the present work we have used BR. 
However, a very recent work~\cite{Jayalath:2011uc} has adopted BF. 
In order to compare the results of this latter work with those of model B, we have investigated the drawback of 
using BF instead of BR in our approach. 
Accordingly, a third model, hereafter called model C, was obtained.

Though we extract simultaneously the partial decay widths for both $S_{11}$ resonances, the
above changes in the data do not affect results for the $S_{11}(1535)$.
In Table~\ref{BR-BF}, results from PDG, Ref.~\cite{Jayalath:2011uc} and our models B and C are given
for $S_{11}(1650)$. The $\chi^2_{d.p.}$ for the three models are comparable, namely,
0.15 (model B), 0.25 (model C) and 0.19 (ref.~\cite{Jayalath:2011uc}).

Model C leads to results in agreement with the two other sets, within the uncertainties therein.
Comparing models B and C, we observe that the most sensitive width is $\Gamma_{S_{11}(1650) \to K \Lambda}$
and to a lesser extent $\Gamma_{S_{11}(1650) \to \eta N}$, while $\Gamma_{S_{11}(1650) \to \pi N}$
increases very slightly.

In Table~\ref{cci-Gam}, results for coupling constant from models B and C are reported. 
We find of cours the same features as for partial decay widths. 
In addition, given the associated uncertainties, it turns out that 
$\Gamma_{S_{11}(1650) \to \eta^\prime N}$ and $\Gamma_{S_{11}(1650) \to K \Sigma}$ change very slightly 
within the two models.
%
\begin{table}[t]
\caption{\footnotesize Strong decay widths (in MeV) for $S_{11}(1650)$.}
\begin{tabular}{ccccccccccc}
\hline \hline
 $\Gamma_{tot}$ && $\pi N$ && $\eta N$&& $K \Lambda$ && Approach && Ref.   \\
\hline
 165 $\pm$ 20 &&128 $\pm$ 29&& 3.8 $\pm$ 3.6 && 4.8 $\pm$  0.7 && {\bf BR} && PDG~\cite{Nakamura:2010zzi} \\
              &&143 $\pm$ 5 && 4.5 $\pm$ 3.0 && 4.8 $\pm$ 0.7  &&  Model B &&  Present work \\
              &&128 $\pm$ 29&& 10.7 $\pm$ 5.8 && 11.5 $\pm$  6.6 && {\bf BF} && PDG~\cite{Nakamura:2010zzi} \\
              &&148 $\pm$ 8 && 9.7 $\pm$ 6.7 && 7.9 $\pm$ 0.3  &&  Model C &&  Present work \\
              &&133 $\pm$ 33 && 12.5 $\pm$ 11.0 && 11.5 $\pm$ 6.4  &&  $1/N_C$-NLO && Jayalath {\it et al.}~\cite{Jayalath:2011uc} \\
\hline \hline
\end{tabular}
\label{BR-BF}
\end{table}
%
\begin{table}[ht!]
\caption{\footnotesize $S_{11}(1650)$-meson-baryon coupling constants ($g_{S_{11} MB}$) in isospin basis.}
\begin{tabular}{cccccccccccccc}
\hline \hline
$\pi N$   &&      $\eta N$    &&    $K \Lambda$   && $K \Sigma$ && $\eta^\prime N$ && Approach && Ref.   \\
\hline
-1.64 $\pm$ 0.03 && 0.35 $\pm$ 0.14 &&  0.53 $\pm$ 0.04 && -2.66 $\pm$ 0.06 && -1.62 $\pm$ 0.14 &&  Model B &&  Present work \\
-1.66 $\pm$ 0.05 && 0.55 $\pm$ 0.16 &&  0.62 $\pm$ 0.09 && -2.49 $\pm$ 0.16 && -1.74 $\pm$ 0.24 &&  Model C &&  Present work \\
\hline \hline
\end{tabular}
\label{cci-Gam}
\end{table}
\begin{table}[ht!]
\caption{\footnotesize $S_{11}(1650)$-meson-baryon coupling constants ($g_{S_{11} MB}$) in particle basis.}

\begin{tabular}{ccccccccc}
\hline \hline
 $\pi^{0}p$ & $\pi^{+}n$ & $\eta p$   & $K^{+}\Lambda$  & $K^{0}\Sigma^{+}$ & $K^{+}\Sigma^{0}$  & $\eta^{\prime}p$  & Ref.   \\
\hline
-0.94 $\pm$ 0.02     & 1.33 $\pm$ 0.03 &  0.35  $\pm$ 0.14 & 0.51 $\pm$ 0.03
               & -2.17 $\pm$ 0.05    &1.53 $\pm$ 0.04 &  -1.62 $\pm$ 0.14      &Model B\\

 -0.96 $\pm$ 0.03     & 1.36 $\pm$ 0.04 &  0.55  $\pm$ 0.16 & 0.62 $\pm$ 0.09
               & -2.03 $\pm$ 0.13    &1.44 $\pm$ 0.09 &  -1.74 $\pm$ 0.24      &Model C\\
\hline \hline
\end{tabular}
\label{cc-c}
\end{table}

Those trends are also present in the coupling constants given in particle basis (Table~\ref{cc-c}).

Taking into account the associated uncertainties to the coupling constants, model C does not significantly
modify the coupling constants ordering obtained in sec.~\ref{2cc} for model B.

To end this section, we give the phase space defined by model C:
\begin{equation}
24.7^\circ \leq \theta_{S} \leq 30.0^\circ~;~19.8\% \leq P_{5q} \leq 31\%~;~3.0\% \leq P^\prime_{5q} \leq 12.6\%\,.
\label{Ranges-c}
\end{equation}

Compared to model B, Eq.~(\ref{Ranges}), the ranges for $\theta_{S}$ and $P_{5q}$ get slightly increased. 
The most significant change concerns $P^\prime_{5q}$, which goes from $11\% \leq P^\prime_{5q} \leq 18$
down to $3\% \leq P^\prime_{5q} \leq 13$. This feature shows the sensitivity of 
$\Gamma_{S_{11}(1650) \to K \Lambda}$ and, to a lesser extent, that of
$\Gamma_{S_{11}(1650) \to \eta N}$ to the five-quark components in $S_{11}(1650)$.
%
%
\section{Summary and Conclusions}
\label{con}
Within a constituent quark approach, we studied the properties of four low-lying 
baryon resonances with respect to their partial decay widths to seven meson-baryon channels
and associated resonance-meson-baryon coupling constants.

The starting point was the  simplest chiral constituent quark model ($\chi$CQM).
The second step consisted in introducing $SU(6) \otimes O(3)$ breaking effects.
Finally, five-quark components in the $S_{11}$ resonances were implemented and investigated.

The outcome of the present work is reported below, focusing on the considered 
nucleon resonances ($S_{11}(1535)$, $S_{11}(1650)$, $D_{13}(1520)$ and $D_{13}(1700)$)
and their strong decays to $\pi N$, $\eta N$, $\eta^\prime N$, $K \Lambda$ and $K \Sigma$
final states.

Within the $\chi$CQM, the only adjustable parameter ($\omega_3$) did not allow reproducing the 
partial widths of resonances. Introducing $SU(6) \otimes O(3)$ breaking, via configuration 
mixing angles $\theta_S$ and $\theta_D$, brought in significant improvements with respect to
the decay widths of the $D_{13}$ resonances, but missed the data for
the $S_{11}$ resonances partial decay widths. 
Nevertheless, this second step allowed fixing the value of $\omega_3$
and extracting ranges for the mixing angles, treated as free parameters. Trying to cure this 
unsatisfactory situation, possible roles due to five-quark component in the baryons' wave
functions were investigated. Given that the latter issue is irrelevant with respect to the 
$D_{13}$ resonances and the properties of which were well descried in the second step, the
final phase of our study was devoted to the $S_{11}$ resonances.

We calculated the partial decay widths $S_{11}(1535)\to\pi N$, $\eta N$ and
$S_{11}(1650)\to \pi $, $\eta N$, $K \Lambda$ in the whole phase space defined by the mixing angle
$\theta_S$ and the probability of five-quark components in each of the two resonances.
Regions of the phase space allowing to reproduce the data for those widths were selected.
Accordingly, that procedure allowed us extracting ranges for partial widths, with decay threshold 
below the relevant resonance mass, and resonance-meson-baryon coupling constants
for the following meson-baryon combinations:
$\pi^{0}p$, $\pi^{+}n$, $\eta p$, $K^{+}\Lambda$, $K^{0}\Sigma^{+}$, $K^{+}\Sigma^{0}$ and $\eta^{\prime}p$.

The main findings of the present work are summarized below with respect to the approaches studied in describing
the properties of the four low-lying nucleon resonances.

\begin{itemize}
  \item The chiral constituent quark approach in three-quark configuration and exact $SU(6)\otimes O(3)$ symmetry 
  is not appropriate to reproduce the known partial decay widths. 
  \item Introducing symmetry breaking effects due to one-gluon-exchange mechanism, allows accounting for the partial 
  decay width of the $D_{13}(1520)$ and $D_{13}(1700)$ resonances, but not for those of $S_{11}$ resonances.
  \item Complementing the formalism with five-quark components in the $S_{11}$ resonances leads to satisfactory results
  with respect to all known partial decay widths investigated here.
  \item The complete formalism puts ranges on the three adjustable parameters, namely, the mixing angle 
  between configurations $|N^{2}_{8}P_{M}\rangle$ and $|N^{4}_{8}P_{M}\rangle$, and five-quark component probabilities 
  in $S_{11}(1535)$ and  $S_{11}(1650)$ resonances.
  \item For $S_{11}(1535)$, the most sensitive entities to the five-quark component turn out to be
  $\Gamma_{S_{11}(1535) \to \eta N}$, $g_{S_{11}K^{+}\Sigma^{0}}$, $g_{S_{11}K^{0}\Sigma^{+}}$ and 
  $g_{S_{11}\eta p}$, all with sizeable magnitudes.
  \item For $S_{11}(1650)$, the same trends as for $S_{11}(1535)$ are observed. 
  In addition $\Gamma_{S_{11}(1650) \to \pi N}$ undergoes significant change due to five-quark mixture.  
  Here, $\eta N$ channel have smaller width and coupling constant compared to the $S_{11}(1535)$ case.
\end{itemize}

To go further, interpretation of recent data, obtained using electromagnetic and/or hadronic probes, within
approaches with reasonable number of free parameters is very desirable. Within the present extended $\chi$CQM approach, 
analysis of the $\gamma p \to \eta p$ data is underway~\cite{next}.
%
%
\begin{acknowledgments}

One of us (C.~S.~A.) thanks X.~H.~Liu and J.~J.~Xie for very helpful discussions.

\end{acknowledgments}

\begin{appendix}
%
%

\section{{\boldmath $S_{11}(1535)$} and {\boldmath $S_{11}(1650)$} resonances mixing angle in 
one-gluon-exchange and one-boson-exchange models}
\label{apdx:mix}

The mixing angle $\theta_{S}$ can be obtained by diagonalizing the following matrix:
\begin{eqnarray}\pmatrix{ \langle
N(^{2}_{8}P_{M})_{\frac{1}{2}^{-}},S_{z}|H_{hyp}|N(^{2}_{8}P_{M})_{\frac{1}{2}^{-}},S_{z}\rangle,
&  \langle
N(^{2}_{8}P_{M})_{\frac{1}{2}^{-}},S_{z}|H_{hyp}|N(^{4}_{8}P_{M})_{\frac{1}{2}^{-}},S_{z}\rangle
 \cr
\langle
N(^{4}_{8}P_{M})_{\frac{1}{2}^{-}},S_{z}|H_{hyp}|N(^{2}_{8}P_{M})_{\frac{1}{2}^{-}},S_{z}\rangle,
& \langle
N(^{4}_{8}P_{M})_{\frac{1}{2}^{-}},S_{z}|h_{hyp}|N(^{4}_{8}P_{M})_{\frac{1}{2}^{-}},S_{z}\rangle\cr}\,,
\end{eqnarray}
where $H_{hyp}$ is the hyperfine interaction between the quarks. 
In the OGE~\cite{De Rujula:1975ge} and OBE models~\cite{Glozman:1995fu}, the explicit forms of $H_{hyp}$ are
\begin{eqnarray}
 H_{hyp}^{OGE}&=&\sum_{i<j}\frac{2\alpha_{s}}{3m_{i}m_{j}}\{
\frac{8\pi}{3}\vec{S}_{i}\cdot\vec{S}_{j}\delta^{3}(\vec{r}_{ij})
+\frac{1}{r_{ij}^{3}}[\frac{3\vec{S}_{i}\cdot\vec{r}_{ij}\vec{S}_{j}\cdot\vec{r}_{ij}}{r_{ij}^{2}}-
\vec{S}_{i}\cdot\vec{S}_{j}]\}\\
H_{hyp}^{OBE}&=&\sum_{i<j}\sum_{F}\frac{g^{2}}{4\pi}\frac{1}{12m_{i}m_{j}}
\vec{\lambda}_{i}^{F}\cdot\vec{\lambda}_{j}^{F}\{[\vec{\sigma}_{i}\cdot\vec{\sigma}_{j}(\frac{\mu^{2}e^{-\mu r_{ij}}}{r_{ij}}-4\pi\delta(\vec{r}_{ij}))]\nonumber\\
&&+
(\frac{3\vec{\sigma}_{i}\cdot\vec{r}_{ij}\vec{\sigma}_{j}\cdot\vec{r}_{ij}}{r_{ij}^{2}}-\vec{\sigma}_{i}\cdot\vec{\sigma}_{j})
\frac{\mu^{2}e^{-\mu r_{ij}}}{r_{ij}}(1+\frac{3}{\mu
r_{ij}}+\frac{3}{\mu^{2}r_{ij}^{2}})\}
\end{eqnarray}
\subsection{One-Gluon-Exchange (OGE) model}
The OGE hyperfine interaction leads to the following matrix elements:
\begin{eqnarray}
\langle
N(^{2}_{8}P_{M})_{\frac{1}{2}^{-}},S_{z}|H_{hyp}^{OGE}|N(^{2}_{8}P_{M})_{\frac{1}{2}^{-}},S_{z}\rangle&=&-C,\\
\langle
N(^{2}_{8}P_{M})_{\frac{1}{2}^{-}},S_{z}|H_{hyp}^{OGE}|N(^{2}_{8}P_{M})_{\frac{1}{2}^{-}},S_{z}\rangle&=&C,\\
\langle
N(^{4}_{8}P_{M})_{\frac{1}{2}^{-}},S_{z}|H_{hyp}^{OGE}|N(^{2}_{8}P_{M})_{\frac{1}{2}^{-}},S_{z}\rangle&=&C,\\
\langle
N(^{4}_{8}P_{M})_{\frac{1}{2}^{-}},S_{z}|H_{hyp}^{OGE}|N(^{4}_{8}P_{M})_{\frac{1}{2}^{-}},S_{z}\rangle&=&0\,,
\end{eqnarray}
with the constant $C=\frac{2\alpha_{s}}{m^{2}}\omega_{3}^{3}\pi^{-\frac{1}{2}}$, where $m$ and $\omega_{3}$  
are the light quark mass and the harmonic oscillator parameter, respectively. 
Then, we obtain $\theta_{S}^{OGE}\simeq32$\textdegree. 

Here a comment is in order with respect to the sign of $\theta_{S}$.
As, reported in Ref.~\cite{Saghai:2009zz}, a non ambiguous entity with respect to that sign is the following ratio:
\begin{eqnarray}\label{eq:MixR}
{\cal {R}} =  \frac {<N|H_m|N(^4P_M)_{{\frac 12}^-}>}
{<N|H_m|N(^2P_M)_{{\frac 12}^-}>},
\end{eqnarray}
with $H_m$ the pseudovector couplings at the tree level.
The ratio ${\cal {R}}$ is a constant determined by $SU(6)\otimes O(3)$ symmetry. 

Notice that in the present work, we have adopted the convention introduced by Koniuk and Isgur~\cite{Koniuk:1979vy}, 
where wave functions are in line with the SU(3) conventions of de Swart~\cite{de Swart:1963gc}. 
In this frame, the constant ${\cal {R}}$ gets a negative value, 
and the relevant mixing angle for the $S-$wave, $\theta_{S}$, turns out positive. 
However, in line with the Hey, Litchfield, and Cashmore~\cite{Hey:1974nc} analysis, 
Isgur and Karl in their early works~\cite{Isgur:1977ef,Isgur:1978xi,Isgur:1978xj,Isgur:1978wd} 
used another convention, for which ${\cal {R}}$ = +1 and $\theta_{S} <$~0. 
In the literature both conventions are being used, often without explicit mention of the utilized convention.

\subsection{One-Boson-Exchange (OBE) model}
The OBE hyperfine interaction results in
\begin{eqnarray}
\langle
N(^{2}_{8}P_{M})_{\frac{1}{2}^{-}},S_{z}|H_{hyp}^{OBE}|N(^{2}_{8}P_{M})_{\frac{1}{2}^{-}},S_{z}\rangle&=&5V_{11}-7V_{00},\\
\langle
N(^{2}_{8}P_{M})_{\frac{1}{2}^{-}},S_{z}|H_{hyp}^{OBE}|N(^{2}_{8}P_{M})_{\frac{1}{2}^{-}},S_{z}\rangle&=&-8T_{11},\\
\langle
N(^{4}_{8}P_{M})_{\frac{1}{2}^{-}},S_{z}|H_{hyp}^{OBE}|N(^{2}_{8}P_{M})_{\frac{1}{2}^{-}},S_{z}\rangle&=&-8T_{11},\\
\langle
N(^{4}_{8}P_{M})_{\frac{1}{2}^{-}},S_{z}|H_{hyp}^{OBE}|N(^{4}_{8}P_{M})_{\frac{1}{2}^{-}},S_{z}\rangle&=&4V_{11}-2V_{00}+8T_{11}\,,
\end{eqnarray}
where $V_{00}$, $V_{11}$ and $T_{11}$ are constants from the orbital integral
\begin{eqnarray}
V_{00}&=&\langle\varphi_{00}|\frac{g^{2}}{4\pi}\frac{1}{12m_{i}m_{j}}
(\frac{\mu^{2}e^{-\mu
r_{ij}}}{r_{ij}}-4\pi\delta(\vec{r}_{ij}))|\varphi_{00}\rangle,\\
V_{11}&=&\langle\varphi_{1m}|\frac{g^{2}}{4\pi}\frac{1}{12m_{i}m_{j}}
(\frac{\mu^{2}e^{-\mu
r_{ij}}}{r_{ij}}-4\pi\delta(\vec{r}_{ij}))|\varphi_{1m}\rangle,\\
T_{11}&=&\langle\varphi_{1m}|\frac{g^{2}}{4\pi}\frac{1}{12m_{i}m_{j}}
\frac{\mu^{2}e^{-\mu r_{ij}}}{r_{ij}}(1+\frac{3}{\mu
r_{ij}}+\frac{3}{\mu^{2}r_{ij}^{2}})|\varphi_{1m}\rangle.
\end{eqnarray}
Taking the same values for the parameters as in
Ref.~\cite{Glozman:1995fu}, we obtain $\theta_{S}=-13$\textdegree. 
However, if one considers contributions from the vector meson exchanges, the absolute value of 
$\theta_{S}$ might be decreased, or even the sign might change~\cite{Glozman:1999ms,Glozman:1998wk}.

Relevance of the OGE versus the OBE has been studied by several authors, see e.g. 
Refs.~\cite{Capstick:2004tb,Yoshimoto:1999dr,He:2003vi,Liu:2005wg}, favoring OGE mechanism, 
endorsed by the present work, as the origin of the $SU(6)\otimes O(3)$ symmetry breakdown.
%
\end{appendix}


%
%
\end{document}